\documentclass[reprint,amsmath,amssymb,aps,]{revtex4-2}

\usepackage{graphicx}% Include figure files
\usepackage{dcolumn}% Align table columns on decimal point
\usepackage{bm}% bold math
\usepackage{gensymb}

\usepackage{color}

\begin{document}

\preprint{APS/123-QED}

\title{Liquid-liquid phase separation driven by charge heterogeneity}

\author{Daniele Notarmuzi}
\email{daniele.notarmuzi@tuwien.ac.at}
\affiliation{Institut für Theoretische Physik, TU Wien, Wiedner Hauptstraße 8-10, A-1040 Wien, Austria}
\author{Emanuela Bianchi}
\affiliation{Institut für Theoretische Physik, TU Wien, Wiedner Hauptstraße 8-10, A-1040 Wien, Austria and CNR-ISC, Uos Sapienza, Piazzale A. Moro 2, 00185 Roma, Italy}

\begin{abstract}
Globular proteins as well as recently synthesized colloids engineered with differently charged surface regions have in common a reduced  bonding valence and a complex interaction pattern dominated by like-charge attraction and opposite-charge repulsion. While the impact of low functionality on the condensation of the liquid phase has been extensively studied, the combined effect of limited bonding valence and particle charge heterogeneity on the liquid-liquid phase separation has not been investigated yet. We numerically tackle this challenge in a systematic fashion by taking advantage of an efficient coarse-grained model grounded into a robust mean-field description. We consider a relatively simple surface pattern consisting of two charged polar caps and an oppositely charged equatorial belt and investigate how the interplay between geometry and electrostatics affect the critical point parameters. We find that electrostatics has a dramatic effect on the condensation of the liquid phase --  especially in the regime of large polar caps. 
\end{abstract}

\maketitle
Globular proteins are compact biomolecules whose effective interactions through the surrounding electrolyte result from the complex interplay of many factors, such as electrostatic forces, specific interactions, hydrogen bonding and hydrophobic effects~\cite{protein_book_2007}. As such, developing realistic models for protein–protein interactions is a challenging task and the coarse-grain level of the description must target to the phenomena of interest~\cite{McManus2016The}.

Within the aim of understanding the liquid-liquid phase separation (LLPS), i.e., the separation into a protein-rich and a protein-poor phase, isotropically interacting colloid models have shed light on how the range of the effective interaction affects the stability of the liquid phase. The predictive power of isotropically, short-ranged attractive models was assessed, e.g., in lysozyme and $\gamma$-crystalline systems~\cite{Benedek_1999,Schurtenberger_2011,Egelhaaf_2015}: for extremely short-ranged attractions, the condensation of the liquid phase was observed to be metastable with respect to the fluid-solid transition~\cite{HagenFrenkel_1994,MillerFrenkel_2004,PaganGunton_2005} and the metastability of the critical point was related to the proteins crystallization mechanism~\cite{tenWoldeFrenkel_1997,VliegenthartLekkerkerker_2000}. The micron-scale counterpart of such a phenomenon has been investigated for instance in systems of colloidal particles interacting via depletion interactions~\cite{deHoog_2001,AndersonLekkerkerker_2002}. 

A successive step forward in the understanding of the LLPS in solutions of globular proteins has been made by acknowledging that protein-protein interactions are intrinsically anisotropic and characterized by a limited bonding valence: coarse-grained models consisting of overall repulsive spheres with a limited number of functional bonding sites -- mimicking the multi-valency of the proteins binding groups -- have shown that on reducing the average valence of the systems the drive for the condensation of the liquid phase is drastically reduced~\cite{Sear_1999,KernFrenkel_2003, Bianchi2006Phase,Espinoza2020Liquid}. The concept of reduced bonding functionality as a control parameter of the LLPS in protein systems is nowadays often used to gain insight into  experimental data~\cite{roosenrunge2014sr,Sweeney_2017,Levy_2020}. The colloidal counterpart of these systems are micron-scale particles with suitably designed surface patterns, often referred to as patchy colloids~\cite{Glotzer_2004,Kretzschmar_2010,Bianchi_pccp_2011,Pine_2012,Sacanna_2017}, whose assembly scenarios are very rich~\cite{Granick_2011,Bianchi2017Limiting,Pine_2020}. 

It is known, however, that proteins are heterogeneously charged~\cite{yang2023jpcb,nakamura1985nature,Li2015Charge,lund2016anisotropic,Bozic2017pH,boubeta2018langmuir,Kosovan_2022} and that their surface charge distribution is dictated by the protonation state of the acid and basic groups along their backbone, which in turn depends on the ionic strength of the solution as well as on the local physio-chemical properties of the environment~\cite{Bozic2017pH,Kosovan_2022}: for a given configuration of ionizable groups, the charge surface pattern and the net particle charge of the protein depend, e.g., on the local pH. Bonds between oppositely charged residues -- often referred to as salt-bridges -- are recognized to play a role in the effective interactions between distinct proteins 
%-- at least in the regime of low salt concentrations, when hydrophobic interactions are too short ranged to play a major role.
and examples of electrostatically driven phase separation have been reported in the literature~\cite{broide1996using,muschol1997liquid,grigsby2001cloud,gogelein2012effect,Zhang2012Charge,lund2012jpcl,roosenrunge2014sr}
At the colloidal level, micron-sized colloids with heterogeneous surface charge distributions have been synthesised with relatively simple surface patterns by means of different approaches~\cite{vanostrum2015jpcm,Zimmermann_2018,Mehr2019sm,Shanmugathasan2022Silica,Virk2023Synthesis} and the electrostatically-driven directional interactions between these complex units have been linked to diverse assembly behaviours, resulting in a broad spectrum of ordered and disordered phases~\cite{sabapathy2017pccp,Kimura_2019,Naderi2020Self,Lebdioua2021Study,Cruz_2016,Blanco_2016,silvanonanoscale,abrikosov2017steering,Ferreira_2017,Cerbelaud_2019,Swan_2019,Vashisth_2021,mani2021stabilizing}.
Colloids engineered with heterogeneously charged surfaces and globular proteins have in common a limited  bonding valence and orientation-dependent effective interactions, where both directional attraction and directional repulsion are present. As these features are crucial both in materials design and in biological applications, several coarse-grained models have been recently proposed in the literature~\cite{hoffmann2004molphys,boon2010jpcm,Bianchi2011Inverse,degraaf2012jcp,yigit2015jcp,hieronimus2016jcp,Brunk_2020,mathews2022molsim,popov2023jpcb}. 
Nonetheless, to the best of our knowledge, model systems with both limited bonding valence and surface charge heterogeneity have not been yet investigated in the contest of the LLPS. We numerically tackle the combined effect of these two features in a systematic fashion by taking advantage of  a relatively simple coarse-grained model grounded into a robust mean-field approach~\cite{Bianchi2011Inverse,bianchi:2015}. 

\begin{figure*}[ht!]
\begin{center}
\includegraphics[width=\textwidth]{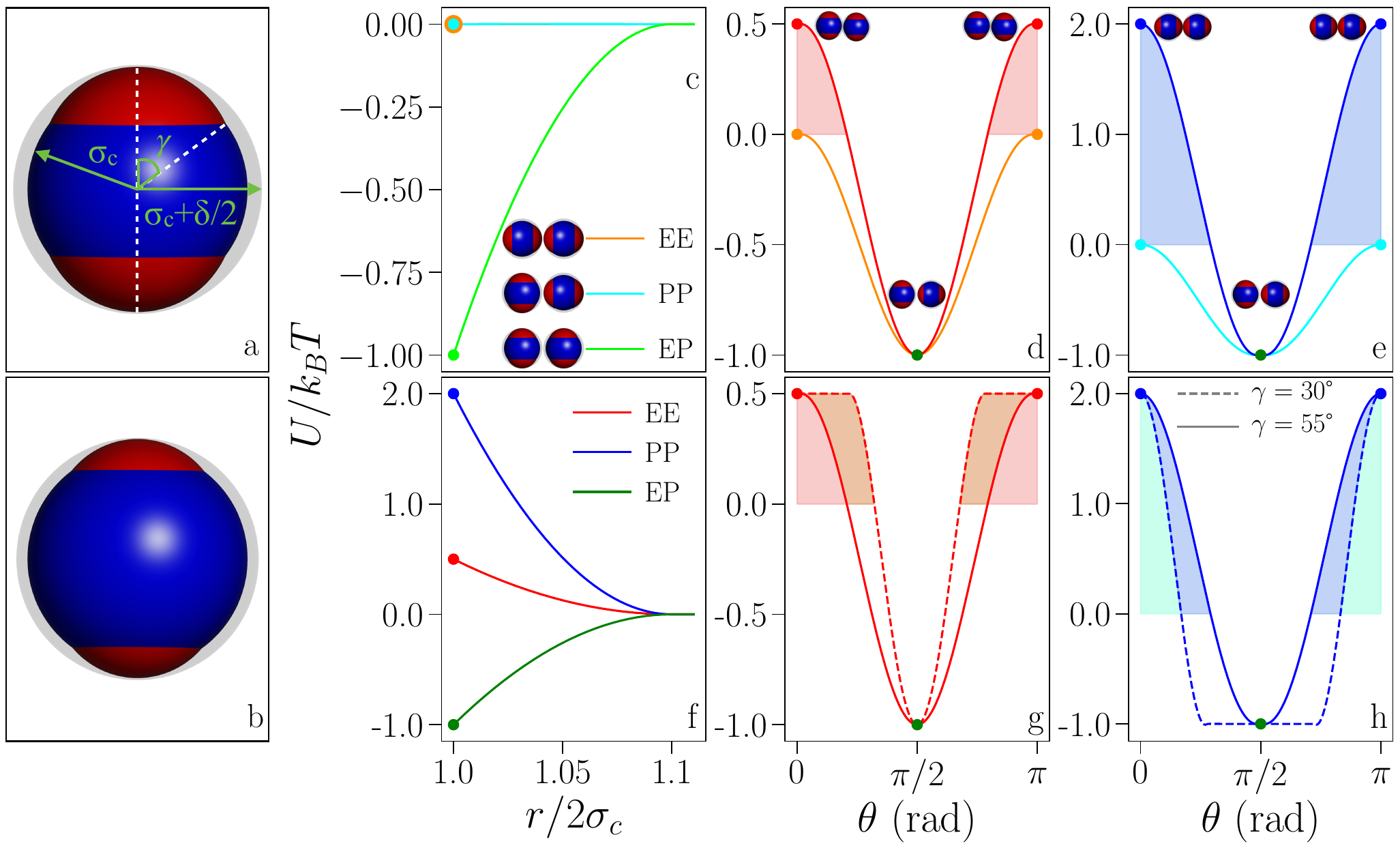}    
\end{center}
\caption{Panels (a) and (b): Inverse Patchy Particle (IPP) model representations: the blue sphere is the hard core particle, the red polar caps depict those portions of the interaction spheres associated to the off-center sites that protrude out of the particle, the gray shade represents the interaction sphere associated to the central site; note that IPPs are a spherical units. The geometrical parameters of the IPP model are highlighted in panel (a): the particle radius, $\sigma_c=0.5$, the particle interaction radius, $\sigma_c+\delta/2$, and the half-opening angle, $\gamma$, as labeled; in (a) $\gamma=55$°, while in (b) $\gamma=30$°. Panel (c): Interaction energy as a function of the inter-particle separation for pairs of particles in three reference configurations -- labeled as EE, PP and EP and shown on the side of the legend -- for IPPs with $u_{\rm EE}=u_{\rm PP} = 0.0$ (and any $\gamma$); note that the orange and cyan lines coincide. Panel (d): Interaction energy at contact ($r=2\sigma_c$) on rotating the particle symmetry axis from the EE, to the EP and back to EE configuration for $\gamma=55$°. The orange line is for $u_{\rm EE}=0.0$, the red line is for $u_{\rm EE}=0.5$. The shaded area spans configurations with $0<U/k_BT<0.5$.
(e) Same as in (d) for the rotation from the PP, to the EP back to the PP configuration. The cyan line is for $u_{\rm PP}=0$, the blue line is for $u_{\rm PP}=2$. The shaded area spans configurations with $0<U/k_BT<2$. (f) Same as in (c) but for IPPs with $u_{\rm EE}=0.5$ and $u_{PP}=2$ (and any $\gamma$). 
(g) Same as in (d) where the red solid line is the same as in (c) (i.e.,  $\gamma=55$° and $u_{\rm EE}=0.5$), while the red dotted line is for $\gamma=30$° and $u_{\rm EE}=0.5$. (h) Same as in (e), where the blue solid line is the same as in (e) (i.e.,  $\gamma=55$° and $u_{\rm PP}=2$), while the blue dotted line is for $\gamma=30$° and $u_{\rm PP}=2$. Colored regions in panels (g) and (h) highlight the difference in energy between equivalent particle configurations -- large versus small patch cases. }
\label{fig:model}
\end{figure*}

In this letter, we focus on a triblock surface charge distribution consisting in a negatively charged equatorial belt and two positively charged, identical polar regions, also referred to as patches  (see panels (a) and (b) of Fig.~\ref{fig:model}). We model the particles as impenetrable hard spheres of radius $\sigma_c=0.5$ (i.e., $2\sigma_c$ sets the unit of length) and represent the triblock pattern with a set of three interaction sites: one in the particle center -- associated to the equatorial belt -- and two off-center sites -- associated to the polar caps. The model underlies two simplifying assumptions that allow for a systematic exploration of particle geometries and interaction strengths. First, as the patches are assumed to be identical, the off-center sites are both at distance $a < \sigma_c$ from the particle center and placed opposite to each other with respect to it, thus giving rise to an axially symmetric, linear quadrupole. Second, the radius $\sigma_p$ of the interaction spheres associated to the off-center sites is constrained by $\sigma_p + a = \sigma_c + \delta/2$, where $\sigma_c+\delta/2$ is the radius of the interaction sphere associated to the central site and $\delta$ defines the particle interaction range. This constraint -- imposed by the screening conditions of the solvent -- implies that the half opening angle, $\gamma$, describing the patch surface area is $\gamma=\arccos{[(\sigma_c^2 + a^2 - \sigma_p^2) / 2a\sigma_c]}$ (see panel (a) of Fig.~\ref{fig:model}). While we fix $\delta=0.2\sigma_c$, we vary $\gamma$ between $30^{\degree}$ and $55^{\degree}$. 

The specific form of the inter-particle potential is based on the postulate that the different contributions stemming from the attraction/repulsion between like-charged/oppositely-charged regions can all be factorized into a characteristic energy strength and a geometric weight factor, where the first takes into account the distinctive interaction energies between differently charged surfaces and the latter takes into account the distance $r$ and relative orientation $\Omega$ between the interacting particles. Namely, the interaction energy between two particles is infinitely repulsive for inter-particle distances $r < 2\sigma_c$ and zero for $r > 2\sigma_c +\delta$, while it is a weighted sum of energy contributions at intermediate distances:
\begin{equation}\label{eq:potential}
U (r,\Omega)  = \sum_{\alpha\beta} \epsilon_{\alpha\beta}w_{\alpha\beta}(r,\Omega)
\end{equation}
where $\alpha$ and $\beta$ specify the interaction sites ($\alpha$ and $\beta$ can be either the central site or one of the off-center sites), $w_{\alpha\beta}$ takes into account the weight of the $\alpha\beta$ contribution to the total pair energy, and $\epsilon_{\alpha\beta}$ characterizes the energy strength of the $\alpha\beta$ interaction type. For each given pair configuration, the $w_{\alpha\beta}$-values are chosen to be proportional to the overlap volume between all pairs of interaction spheres pertaining to the selected $\alpha\beta$ interaction~\cite{Bianchi2011Inverse}. While the $\epsilon_{\alpha\beta}$-values can be fixed by mapping the coarse-grained model to an orientation-dependent DLVO-like description~\cite{Bianchi2011Inverse,bianchi:2015}, we choose here to change them arbitrarily by fixing the value of $U$ in three reference configurations -- labeled as EE, EP and PP -- and inverting the three by three system resulting from Eq.~\ref{eq:potential} -- see SM for details 
We thus characterize the different systems with the set of contact energies at those configurations $\mathbf{u} = \{u_{\rm EE}, u_{\rm EP}, u_{\rm PP}\}$ -- rather than with the set of characteristic energy strengths $\epsilon_{\alpha\beta}$. The rationale behind our choice is that, at fixed geometric parameters, the $\epsilon_{\alpha\beta}$-values are fully determined by the charges associated to the different surface areas and thus, on changing the net particle charge  (by varying the pH of the solution) the characteristic balance between the equator-equator (EE), equator-pole (EP) and pole-pole (PP) interaction energies can be tuned~\cite{Bianchi2011Inverse,bianchi:2015}. For each patch size, we build a set of systems with different $\mathbf{u}$ to systematically explore the role of the attractive and repulsive directional interactions on the LLPS. Note that we set $u_{EP}=-1.0$ to be the energy scale of all models.
Characteristic representations of the interaction energy are reported in Fig.~\ref{fig:model}: $U/k_BT$ as a function of the inter-particle distance in panels (c) and (f), and as a function of the mutual particle orientation at contact in panels (d)-(e) and (g)-(h) -- for two different energy sets $\mathbf{u} = \{0, -1, 0\}$ and $\mathbf{u} = \{0.5, -1, 2\}$ at the largest ($\gamma=55$°) and the smallest ($\gamma=30$°) patch size. We refer to IPPs with any $\gamma$ and these two energy sets as, respectively, $\rm IPP_{ro}$ (repulsions off) -- as both EE and PP repulsions are artificially set to zero --  and $\rm IPP_{ref}$  (reference) -- as the ratio between the EE and the PP repulsion has been observed in many neutral or slightly charged inverse patchy colloid systems~\cite{Bianchi2011Inverse,bianchi:2015,silvanonanoscale,Noya2014}.

We investigate the behaviour of the selected IPP systems via Grand Canonical Monte Carlo (MC) simulations by adapting the publicly available code published with Ref.~\cite{Rovigatti2018How}. We consider a cubic simulation box of linear size $L=8$ and fix a maximum number of particles $N_{max}$ allowed in it.  
A MC step is defined as $N_{max}$ MC moves, where a move is either a roto-translational move, i.e., the contemporary translation and rotation of a single particle~\cite{Rovigatti2018How}, or the insertion/deletion of a particle, attempted with probability 0.01. Critical points
are identified by 12 parallel runs, each with $5 \cdot 10^7$ MC steps, $2.5 \cdot 10^6$ of which used to equilibrate the system. Histogram reweighting~\cite{Ferrenberg1988New} is then used to match the distribution of $\mathcal{M}=N+sE$ to the Ising magnetization distribution~\cite{Bruce1992Scaling},
where $N$ is the number of particles, $E$ is the total energy of the system and $s$ is a field
mixing parameter~\cite{Bruce1992Scaling} -- see SM for details.

\begin{figure*}[ht!]
\begin{center}
\includegraphics[width=\textwidth]{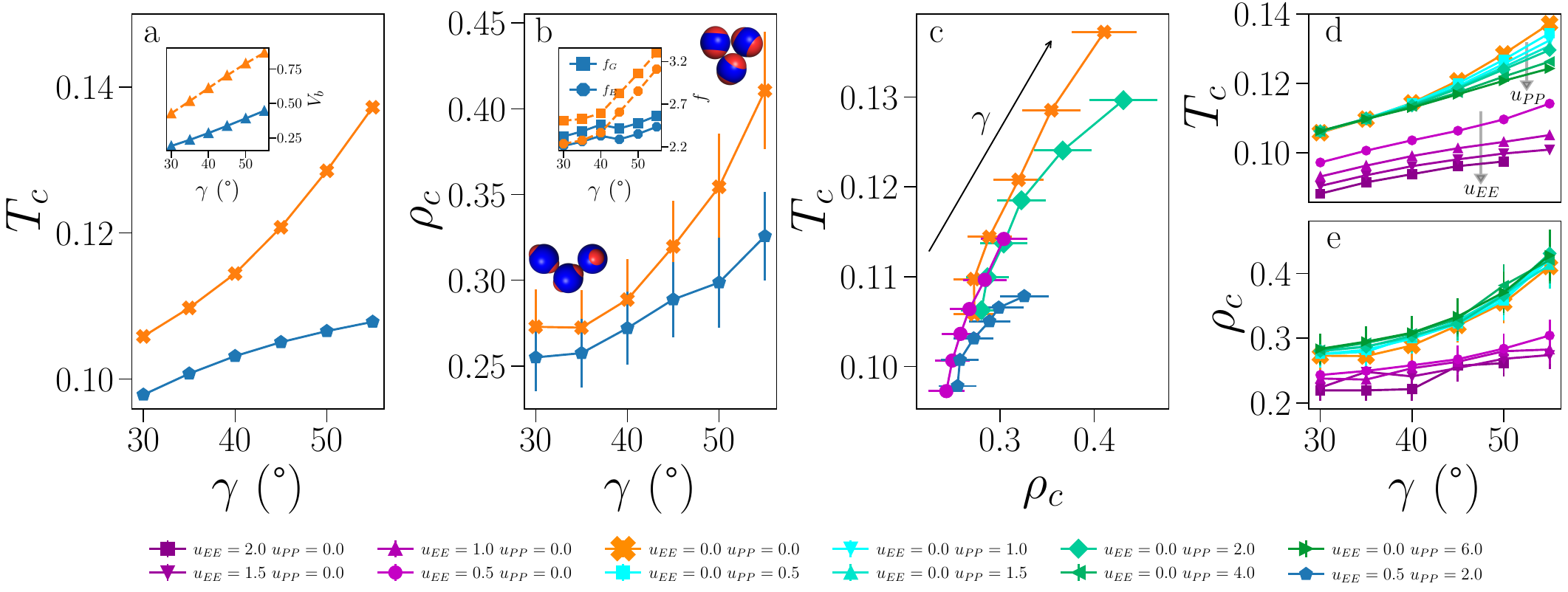}    
\end{center}
\caption{Critical behavior overview: (a) critical temperature, $T_c$, and (b) critical density, $\rho_c$, as functions of $\gamma$, for $\rm IPP_{ro}$ (orange) and $\rm IPP_{ref}$ (blue) systems, as labeled. Inset of panel (a): average particle functionality $f$ as a function of $\gamma$, where bonds are assigned to pairs of particles according either to a geometric ($f_{\rm G}$, squares) or to an energetic ($f_{\rm E}$, circles) criterion; colors as labeled. 
Inset of panel (b): bonding volume $V_b$ as a function of $\gamma$; colors as labeled. Also shown in panel (b) two typical trimers for $\gamma = 30$° (bottom left) and $\gamma = 55$° (upper right), sampled at the critical point of the $\rm IPP_{ro}$ system. (c) Critical point behaviour in the $T-\rho$ phase diagram for $\rm IPP_{ro}$ (orange) and $\rm IPP_{ref}$ (blue) systems, as well as for systems with $u_{EE}=0.0$, $u_{PP}=2.0$ (green) and  $u_{EE}=0.5$, $u_{PP}=0.0$ (magenta). Beyond the $\rm IPP_{ro}$ and $\rm IPP_{ref}$ systems: (d) $T_c$ and (e) $\rho_c$ versus $\gamma$ on increasing either the EE (magenta shades) or the PP (green shades) repulsion with respect to the $\rm IPP_{ro}$ systems, as labeled.}
\label{fig:results}
\end{figure*}

We start our discussion, by considering $\rm IPP_{ro}$ systems as they allow us to isolate the role played by the patch extension $\gamma$. 
As shown in Fig.~\ref{fig:results}, both the critical temperature $T_c$ (panel (a), main) and the critical density $\rho_c$ (panel (b), main)
decrease monotonically as $\gamma$ decreases, thus implying that the LLPS region shrinks on reducing the patch size. 

%% physical interpretation of T_c trend: the bonding volume
Previous studies suggest that the LLPS moves to lower temperatures and densities when the particle bonding valence -- i.e., the maximum number of bonded nearest neighbours -- is reduced by built-in particle features~\cite{Bianchi2006Phase,Foffi2007,Sweeney_2017,Espinoza2020Liquid}. In the contest of conventional patchy colloids, this valence can be quantified by the average number of patches per particle -- if each patch can form only one bond -- or by the percentage of particle surface covered by the patches -- i.e., by the patch size~\cite{Bianchi2006Phase,Foffi2007}.  In contrast, the valence of IPPs cannot be easily related to the number of patches, as the one bond per patch condition is never guaranteed, nor to their surface extension, because of the complementary nature of the interactions: small patches offer little volume to a large equatorial region, while large patches can form bonds only with thin equatorial belts. This conundrum is resolved by directly measuring the amount of volume available to bonding for a given pair of IPPs.

Two particles are defined to form a geometric bond, $G_b$, if they are within their interaction distance $2\sigma_c+\delta$. An energetic bond, $E_b$, is a geometric bond with energy less than zero. The bonding volume, $V_b$, is calculated by evaluating the amount of geometric bonds that are also energetic bonds for any configuration of two interacting particles (see SM for details). We report $V_b$ as a function of $\gamma$ in Fig~\ref{fig:results}b, inset: consistently with the behaviour of $T_c$, it monotonically increases with $\gamma$, thus meaning that the number of energetically bonded configurations of a pair of particles grows continuously with the size of the polar caps even for significantly reduced equatorial regions. 

The behavior of $V_b$ suggests that the average functionality of the particles, $f$, should also monotonically grow with $\gamma$. We calculate $f$ as the average number of bonds per particle in the simulations: again we distinguish geometric and energetic bonds, thus resulting in $f_G$ and $f_E$, respectively -- as reported in Fig.~\ref{fig:results}a, inset. 
Note that in order to evaluate these quantities, the critical point must be known with high precision and simulations must be performed exactly at the inferred critical point (see SM for a detailed discussion). It must also be noted that while the functionality of conventional patchy colloids is often associated to the number of attractive sites and is thus independent on the thermodynamic conditions, in our definition
$f$ is a state-dependent quantity. Consistently with the behavior of $V_b$, both $f_G$ and $f_E$ monotonically grow with $\gamma$, meaning that -- despite the complementarity of polar and equatorial regions -- the average bonding valence of the system always grows with the patch size (at least in the investigated regime of $\gamma$-values), similar to conventional patchy colloids. 

%% physical interpretation of rho_c trend: compact vs disperse structures
The reduction of the particle bonding valence is also accompanied by structural changes in the liquid phase: Fig.~\ref{fig:results}b shows two trimers obtained from samples at the critical point as examples of possible structures observed in the simulations: a branched trimer formed by IPPs with $\gamma=30$° (lower left) and a compact trimer formed by IPPs with $\gamma=55$° (upper right). While for large $\gamma$-values both compact and branched trimers are observed, for small $\gamma$-values, compact trimers are almost entirely absent. These considerations extend beyond the simple case of trimers and can be made quantitative by explicitly measuring the average radius of gyration of the emerging clusters~\cite{Stauffer1992Introduction} (see SM for a more detailed discussion). Structures ranging from timers to large clusters are in general more compact at large $\gamma$, meaning that branched structures -- i.e., less dense clusters -- become increasingly rare for large patch sizes (or, equivalently, for thin equatorial belts). On decreasing $\gamma$ (or, equivalently, on increasing the equatorial belts), less compact structures emerge, $\rho_c$ reduces and, as a consequence, the resulting liquid phase is reminiscent of the ``empty liquid" observed for conventional patchy colloids.

In summary, in absence of directional repulsive interactions, IPPs with small patches have a small bonding volume -- leading to a low $T_c$ -- and an associated reduced functionality, which supports the formation of branched clusters -- leading to a low $\rho_c$.

The role of the electrostatic repulsion can be assessed by considering ${\rm IPP_{ref}}$ systems: in this case, both $u_{EE}$ and $u_{PP}$ are different from zero, with $u_{EE} << u_{PP}$, as observed in many neutral or slightly charged inverse patchy colloid systems. 
We report the corresponding $T_c$ and $\rho_c$ as a function of $\gamma$ in Fig.~\ref{fig:results}, main panel (a) and (b), respectively. The behaviour of both critical parameters is again monotonically decreasing with decreasing $\gamma$, where the effect of the directional electrostatic repulsion is to significantly reduce both $T_c$ and $\rho_c$ with respect to the ${\rm IPP_{ro}}$ systems. This electrostatically-driven shift towards even lower temperatures and densities is $\gamma$-dependent: for small patches, the change with respect to the ${\rm IPP_{ro}}$ counterpart is about $8\%$ in $T_c$ and $12\%$ in $\rho_c$, while for large $\gamma$-values, it raises up to about $27\%$ and $32\%$ respectively.
Consistently with the previous analysis, the bonding volume shown in the inset of Fig.~\ref{fig:results}b remains monotonically growing with $\gamma$, but it decreases
by more than 50\%
with respect to the ${\rm IPP_{ro}}$ systems. Similar to the behaviour of the critical parameters, the electrostatically-driven reduction of $V_b$ is also not constant with $\gamma$ but rather grows with it. The inset of Fig.~\ref{fig:results}a shows that also the functionalities $f_G$ and $f_E$ decrease when electrostatic repulsion is present, by a factor that is strongly $\gamma$-dependent and that is negligible for small patches, while it is particularly large for large patches, mirroring the behaviour of the bonding volume as expected. 

To better visualize the dramatic effect electrostatics has on the liquid-liquid critical point, we display $T_c$ and $\rho_c$ in the $T-\rho$ phase diagram for both ${\rm IPP_{ro}}$ and ${\rm IPP_{ref}}$ systems (see Fig.~\ref{fig:results}c). This representation clearly shows that, beyond a reduction of both $T_c$ and $\rho_c$ when the electrostatic repulsion is on, the range spanned by these two parameters as $\gamma$ grows shrinks, indicating that geometry becomes less crucial when it interplays with electrostatics.  In the same figure, we report the critical point of two additional systems interpolating between the ones already discussed, where $u_{\rm EE}$ and $u_{\rm PP}$ are varied independently. We observe that, while both directional repulsions are able -- alone -- to disfavour the condensation of the liquid phase, the effect of a slight increase of the equatorial repulsion (from zero to 0.5) is more effective than a much larger increase of the polar repulsion (from zero to 2). 

Results obtained by further varying $u_{\rm EE}$ and $u_{\rm PP}$ individually are shown in Fig.~\ref{fig:results}d,e. The critical temperature and density remain monotonically increasing with $\gamma$. The gradual change of parameters highlights that the curvature of $T_c$ as a function of $\gamma$ changes continuously from positive to negative as $u_{\rm EE}$ grows, while it diminishes as $u_{\rm PP}$ grows but does not changes sign. Also, we observe that $T_c$ decreases significantly as $u_{\rm EE}$ grow at any $\gamma$, while the effect of increasing $u_{\rm PP}$ is significant only at large $\gamma$-values while for $\gamma=30$° $T_c$ only slightly grows with $u_{\rm PP}$. The behavior of $T_c$ is in good agreement with the behavior of the bonding volume (see SM): $V_b$ decreases as electrostatic repulsion grows, with $u_{\rm EE}$ having more effect than $u_{\rm PP}$; moreover, its curvature changes sign with increasing $u_{\rm EE}$, while it does not with increasing $u_{\rm PP}$. 
Also for $\rho_c$, the EE repulsion has a more significant effect than the PP repulsion; in this case while $\rho_c$ increases as $u_{\rm EE}$ grows, it (slightly) increases with $u_{\rm PP}$, thus meaning that it is the EE repulsion that mostly leads to the formation of a particularly diluted liquid. 
Notably, the  highest critical temperature is more the 40\% larger than the lowest and the largest $\rho_c$ is almost 80\% larger than the smallest, evidencing the strong effect of the electrostatic repulsion.

In conclusion, in this Letter we investigate the critical behavior of a family of systems characterized by both directional attraction and directional repulsion, mimicking, respectively, the like-charge and opposite-charge contributions that characterize the interaction between heterogeneously charged objects, such as, for instance, proteins. While the LLPS in protein systems has so far been described by means of particle models with a limited bonding valence, in our class of systems the bonding valence is not a built-in feature of the model but rather emerge as a consequence of the complex interaction pattern. Our aim here is to investigate how the geometry of the surface charge distribution and the charge imbalance (or, equivalently, the net particle charge) affect the location of the critical point in the phase diagram. 
For what concerns the effect of the geometry, we find out that the directional attraction emerging from the patchiness of the surface patterns already reduces the particle bonding valence, thus bringing the critical point towards lower temperatures and densities; this is consistent with the ``empty liquid" scenario presented for conventional patchy colloids~\cite{Bianchi2006Phase} and supports the metastabilty of the LLPS in protein systems~\cite{Espinoza2020Liquid}.  As the charge imbalance controls the ratio between the EE and the PP repulsion, we investigate the effects of the net particle charge by independently varying these two repulsions: we note that both the critical parameters are by far more affected by the EE repulsion, rather than by the PP one. This is possibly due to the fact that 
the number of configurations where patch-patch interactions dominate between pairs of IPPs is extremely low; further investigations on the structural properties of the aggregated systems at criticality can elucidate this point~\cite{unpublished}. 
In general, we observe that, on the top of the reduction of the bonding valence induced by the directional attraction, the directional repulsion further and significantly reduces the particle functionality. As the number of bonded pair configurations reduces with the directional repulsion, an electrostatically-driven shift of the critical point to lower temperatures and densities is introduced, thus significantly disfavouring the condensation of the liquid phase.\nocite{Tsypin2000Probability, Sciortino2007Self}

DN and EB acknowledge support from the Austrian Science Fund (FWF) under Proj. No. Y-1163-N27. Computation time at the Vienna Scientific Cluster (VSC) is also gratefully acknowledged.

\appendix

\section*{Supplemental Material}

Numerical simulations of the Inverse Patchy Particle (IPP) model have been performed by adapting the publicly 
available code published with Ref.~\cite{Rovigatti2018How}. The resulting code, together
with data analytics tools to reproduce the results presented in this paper, is
available at~\cite{GitCode}.

\section{Interaction energy}

As stated in the main text, the interaction energy between a pair of IPPs is 
\begin{equation}
U (r,\Omega)  = \sum_{\alpha\beta} \epsilon_{\alpha\beta}w_{\alpha\beta}(r,\Omega)\
\end{equation}
where $\alpha$ and $\beta$ specify the interaction sites ($\alpha$ and $\beta$ can be either the central site or one of the off-center sites), $w_{\alpha\beta}$ takes into account the weight of the $\alpha\beta$ contribution to the total pair energy (and thus depends on $r$ and $\Omega$), and $\epsilon_{\alpha\beta}$ characterizes the energy strength of the $\alpha\beta$ interaction type. The $w_{\alpha\beta}$s are chosen to be proportional to the overlap volume between pairs of interaction spheres pertaining the selected $\alpha\beta$ interaction type for the given particle configuration~\cite{Bianchi2011Inverse}. When the $\epsilon_{\alpha\beta}$-values are fixed by mapping the coarse-grained model to an orientation-dependent DLVO-like description~\cite{Bianchi2011Inverse,bianchi:2015},  they are directly related to the amount of charge located at the interaction sites.
Here, to systematically address the role played by electrostatic interactions, we have varied
the energy strength arbitrarily. To this aim, the vector $\boldsymbol{\epsilon}=(\epsilon_{EE}, \epsilon_{EP}, \epsilon_{PP})$ has been computed
by solving the system
\begin{equation}\label{eq:system}
    W^{-1} \mathbf{u} = \boldsymbol{\epsilon},
\end{equation}
where $W^{-1}$ is the inverse of the matrix whose elements $W_{\alpha \beta}^{AB}$ are the total overlap volumes between all the $\alpha\beta$ sites given the AB configuration. The AB configurations are the reference configurations EE, EP, PP used in the main text and are chosen so that the interaction energy $U$ is dominated by the center/center (cc), center/off-center (co) and off-center/off-center (oo) interaction respectively.
To clarify with an example, $W_{co}^{\rm PP}$ is the sum of the overlap volumes
between the interaction sphere associated to the center of one particle and the interaction spheres associated to both the off-center sites of the other particle, plus the vice-versa, when the mutual orientation between the two particles is specified by the PP configuration.

\section{Grand Canonical Monte Carlo simulations}

The critical points displayed in the main text have been identified by means of Monte Carlo (MC) simulations in the Grand Canonical (GC) ensemble. Observables of a GCMC simulation are the system energy $E$, varying as a consequence of the dynamics, and the particle number $N$, varying as particles can be inserted or removed from the 
simulation box. Simulations were characterized by a maximum number of particles $N_{max}$ allowed in the simulation box.  A MC step was defined as $N_{max}$ MC moves, where the moves used were the insertion/deletion of a particle, attempted with probability 0.01, and a single particle rototranslational (RT) move, i.e., the contemporary  translation and rotation of a single particle~\cite{Rovigatti2018How}. The latter move was attempted with probability 0.99. The maximum translation length (0.05) and maximum rotation angle (0.1) have been set to have an average acceptance rate of the RT move around 0.3  when the model was near the critical point. The average acceptance rate was higher in the diluted phase and lower in the dense phase. The simulation box was a cube with linear size $L=8$.

\section{Identification of the critical point}

To identify the critical point for a given set of $u_{\rm EE}, u_{\rm PP}$ and $\gamma$, numerous short simulations for different values of the temperature $T$ and of the chemical potential $\mu$ have been performed, in order to approximately locate the coexistence region. Once the coexistence region was identified, few different values of $T$ and $\mu$ have been simulated, using 12 independent GCMC simulations per state point.  Each simulation started with $N_0=180$ particles and equilibrated for $2.5 \cdot 10^6$  MC steps, which  was verified a posterior to be a sufficiently large equilibration time for all simulations. The total run time was set to $5 \cdot 10^7$ MC steps per simulation. One value of the observables was collected every $10^3$ MC steps and one
configuration was saved every $5\cdot10^4$ MC steps, resulting in 47500 datapoints and 950 configurations per simulation. This procedure resulted in a total of $57\cdot10^4$ values of $E$ and $N$ per state point and 11400 configurations. 

At the critical point the the probability distribution of the scaling variable $\mathcal{M}=N+sE$ is the same (up to second order corrections that vanish in the thermodynamic limit) of the distribution of the magnetization of the Ising model~\cite{Bruce1992Scaling}, where $s$ is a fitting 
parameter with non universal values. Using data generated simulating the state point $(T,\mu)$ and exploiting the histogram reweighting method~\cite{Ferrenberg1988New}, it is possible to identify new values of $(T', \mu')$, as well as to fit an optimal value of $s$, such that the distribution of $\mathcal{M}$, rescaled to have unit variance, matches the Ising magnetization distribution, computed as in Ref.~\cite{Tsypin2000Probability}. Numerical simulations have been performed for each ($u_{\rm EE}, u_{\rm PP}$, $\gamma$) set until the fit of the reweighted distribution of $\mathcal{M}$ to the Ising magnetization distribution produced an error lower than 0.140, where the error is simply the norm of the difference between the two 
functions. The resulting values of $T'$ and $\mu'$ have been ascribed to be the critical ones.

\begin{figure*}[!htb]
\includegraphics[width=\textwidth]{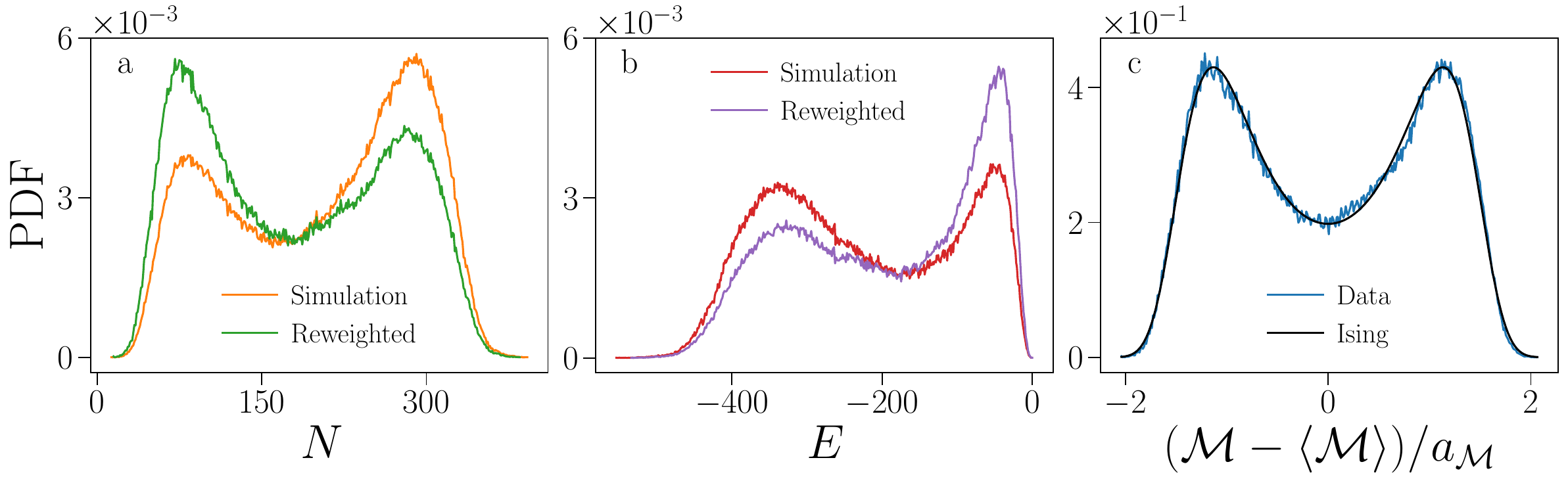}
\caption{Critical distributions before and after reweighting for $u_{\rm EE}=u_{\rm PP}=0.0$ and $\gamma=50$°. (a) The  distribution of the particle number $N$ as obtained from the simulation and after 
reweighting. (b) Same as in (a), but for the energy distribution. (c) The distribution of the scaling variable $\mathcal{M}$, compared to the Ising magnetization distribution. The variable $a_{\mathcal{M}}$ has been chosen for the distribution to have unit variance. 
}
\label{fig:SM_reweight}
\end{figure*}

Fig.~\ref{fig:SM_reweight} exemplifies the result of the procedure described above for $u_{\rm EE}=u_{\rm PP}=0.0$ and $\gamma=50$°. Simulations have been performed at the state point $(T, \mu)= (0.1286, -0.427)$. The figure shows the original distribution of $N$ and $E$ as obtained from the simulations and the reweighted distributions associated to the critical parameters $(T_c, \mu_c)=(0.1285, -0.428)$. The values of $T_c$ and $\mu_c$ have been obtained by matching the distribution of $\mathcal{M}$ to the Ising distribution, as shown in the figure. The fit further provided the optimal value $s=0.25$.

\section{Simulations at the critical point}

The procedure explained above allows to infer the critical parameter $T_c$ and $\mu_c$ without performing a numerical simulation exactly at $T=T_c$ and $\mu=\mu_c$. The histogram reweighting technique further allows to compute the critical density $\rho_c$ and the critical energy density $\epsilon_c$: inferring $T_c$ and $\mu_c$ automatically provides with the joint particle number - energy distribution $P(N, E)$. The latter can be then marginalized to finally obtain $\rho_c = \int_N P(N') dN' / V$ and the critical energy density $u_c = \int_{E} P(E') dE' / V$. However, the analysis of the structural properties of the critical phases requires further simulations to be performed, which must be sufficiently fine tuned to effectively allow for the samples to represent critical behaviour.  

\begin{figure*}[!htb]
\includegraphics[width=\textwidth]{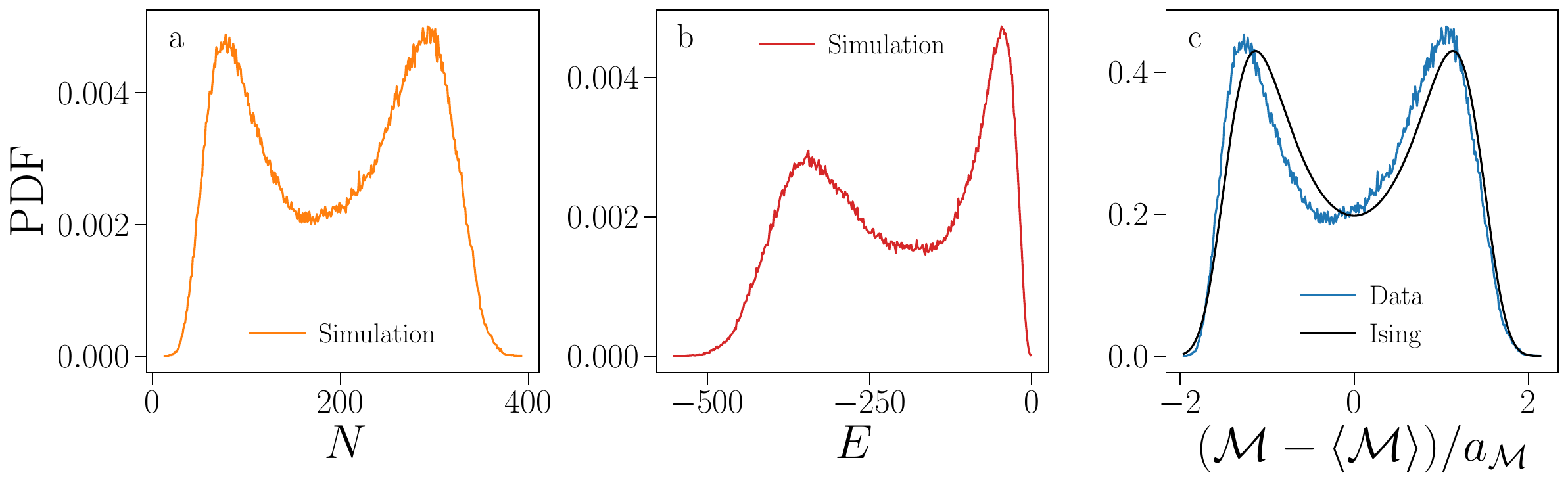}
\caption{Tentatively critical distributions for $u_{\rm EE}=u_{\rm PP}=0.0$ and $\gamma=50$°. (a) The  distribution of the particle number $N$ as obtained from the simulation at $(T_c, \mu_c)=(0.1285, -0.428)$. (b) Same as in (a), but for the energy distribution. (c) The distribution of the scaling variable $\mathcal{M}$, compared to the Ising magnetization distribution. The variable $a_{\mathcal{M}}$ has been chosen for the distribution to have unit variance.
}
\label{fig:SM_fit1}
\end{figure*}

A simple way to verify whether a simulation is sufficiently fine tuned follows from the scaling properties associated to the critical point. In particular, if a simulation is performed at the critical point, then the distribution of the variable $\mathcal{M}$ must coincide with the Ising magnetization distribution without the need of using histogram reweighting. The simulations performed to infer the critical point have  all been made using four digits precision for the temperature and three digits precision for the chemical potential. Hence, it is natural to study the critical configurations by using the same number of digits. As stated, the critical point inferred for $u_{\rm EE}=u_{\rm PP}=0.0$ and $\gamma=50$° is $(T_c, \mu_c)=(0.1285, -0.428)$. Performing numerical simulations with these settings provide with the results shown in Fig.~\ref{fig:SM_fit1},
where the distribution of $N$ and $E$ were used to compute the distribution of $\mathcal{M}$ without reweighting, only allowing for the parameter $s$ to be optimized again. The match between the distribution of $\mathcal{M}$ and the Ising magnetization distribution is clearly not good. It follows that the simulation can not be considered at the critical point and the configurations obtained can not be used to compute the average functionality.

\begin{figure*}[!htb]
\includegraphics[width=\textwidth]{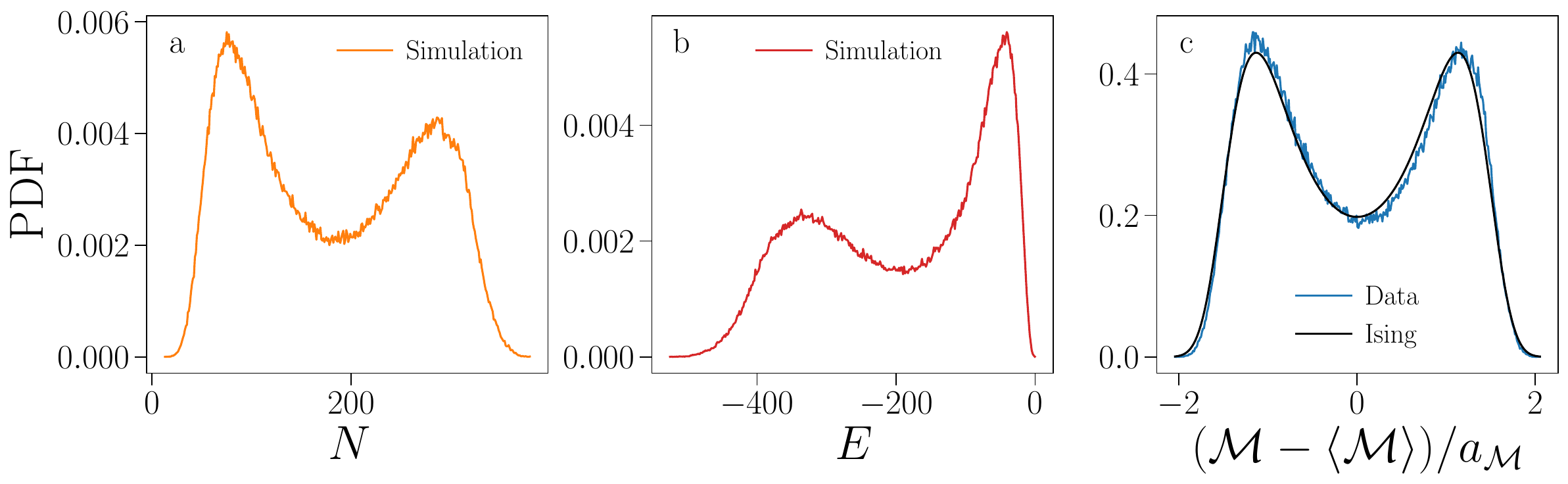}
\caption{Critical distributions for $u_{EE}=u_{PP}=0.0$ and $\gamma=50$°. (a) The  distribution of the particle number $N$ as obtained from the simulation at
$(T, \mu) = (0.128519479875, -0.428066275416)$. (b) Same as in (a), but for the energy distribution. (c) The distribution of the scaling variable $\mathcal{M}$, compared to the Ising magnetization distribution. The variable $a_{\mathcal{M}}$ has been chosen for the distribution to have unit variance.
}
\label{fig:SM_fit2}
\end{figure*}

The reason why the simulations do not reproduce the correct scaling properties of the critical point is that the thermodynamic parameters were not fine tuned enough. The fact that a good fine tuning is required for a system to truly display critical behavior is a natural consequence of the divergence of the susceptibility at the critical point: as the system is extremely susceptible, a small change in the thermodynamic parameters can lead to large changes in the model's behaviour. Hence, the same simulations were performed again, but using more digits, so to better tune the system at criticality. Results for $(T, \mu) = (0.128519479875, -0.428066275416)$ are shown in Fig.~\ref{fig:SM_fit2}. As in the case of Fig.~\ref{fig:SM_fit1}, the variable  $\mathcal{M}$ has been computed without reweighting and only allowing the parameter $s$ to vary, obtaining again $s=0.25$. The configurations sampled in these simulations can be used to compute the average functionality.
As for the determination of critical points, 12 parallel simulations were run, each with run time set to $5 \cdot 10^7$ MC steps per simulation. One value of the observables $N$ and $E$ was collected every $10^3$ MC steps and a configuration was saved every $5\cdot10^4$ MC steps.

\section{Compact vs branched structures}

Compact clusters have smaller radius of gyration if compared to branched clusters of the same size. The radius of gyration of a given cluster of size $s$ can be computed as
\begin{equation}
    R_g = \frac{1}{s} \sum_i | \mathbf{r_i} - \mathbf{r_{cdm}} |^2
\end{equation}
where $\mathbf{r_{cmd}}$ is the center of mass of the cluster. Stauffer and 
Aharony~\cite{Stauffer1992Introduction} define three different ways to average 
$R_g$ over the whole system, i.e.,
\begin{equation}
    \xi_k = \frac{\sum_s \langle R_g^2 \rangle_s s^k n_s}{\sum_s s^k n_s} \hspace{0.5em}{\rm with} \hspace{0.5em} k=0,1,2
    \label{eq:xi_k}
\end{equation}
where $n_s$ is the number of clusters of size $s$ and the symbol $\langle \cdot \rangle_s$ implies that the average is done over clusters of size $s$. 

\begin{figure*}
\includegraphics[width=\textwidth]{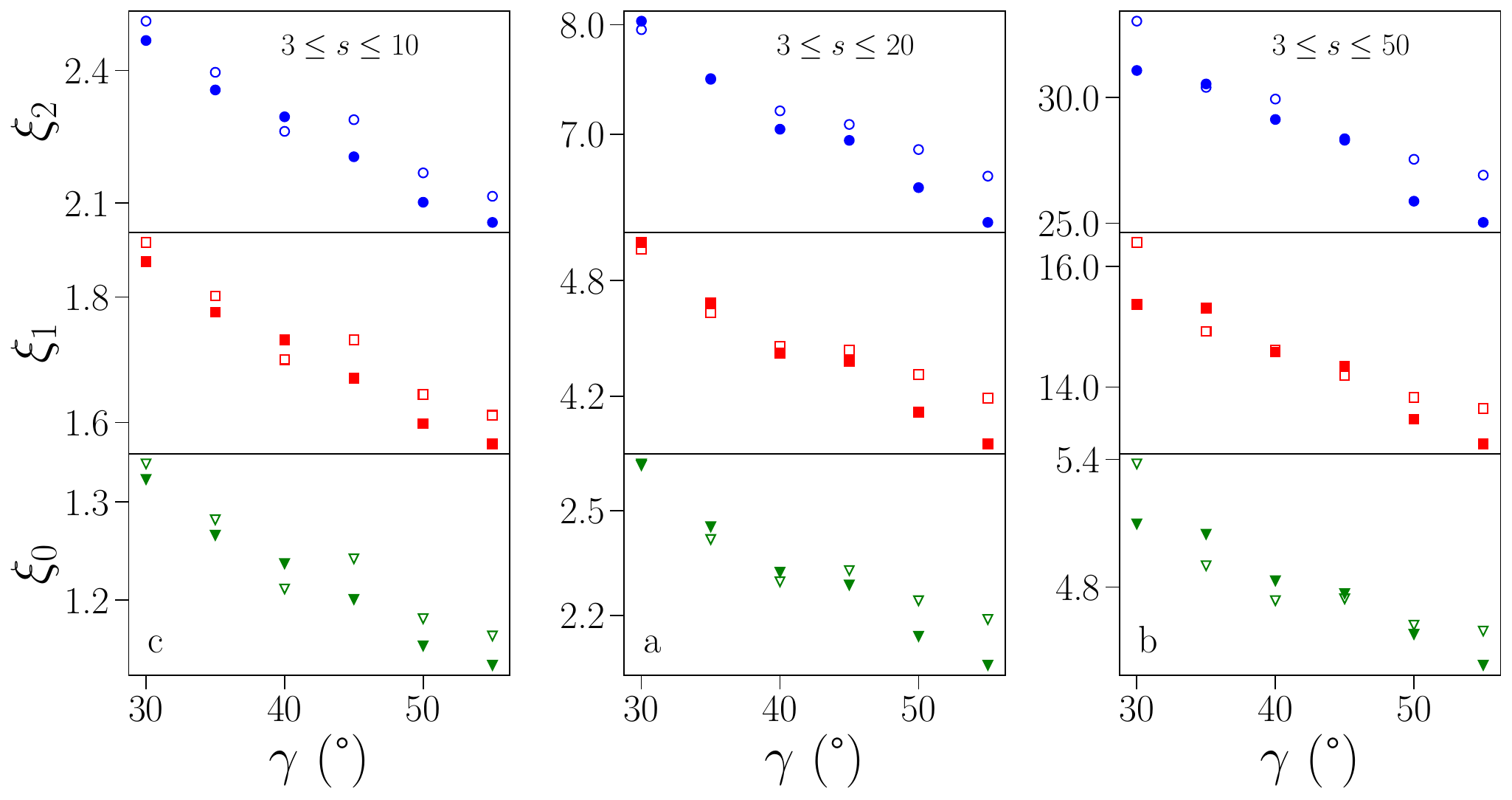}
\caption{System averaged radius of gyration as defined in Eq.~(\ref{eq:xi_k}). Blue circles in the top panels show $\xi_2$, red squares in the middle panels show $\xi_1$ and green triangles in the lower panels show $\xi_0$. Filled markers represent data for  $u_{\rm EE} = u_{\rm PP} = 0.0$ ($\rm IPP_{\rm ro}$), empty markers are for $u_{\rm EE}=0.5$, $u_{\rm PP}=2.0$ ($\rm IPP_{\rm ref}$). (a) The summation in Eq.~(\ref{eq:xi_k}) ranges from $s=3$ to $s=10$. (b) Same as in (a), but the summation in Eq.~(\ref{eq:xi_k}) is from $s=3$ to $s=20$. (c) Same as in (a), but the summation in Eq.~(\ref{eq:xi_k}) is from $s=3$ to $s=50$.
}
\label{fig:xi_k}
\end{figure*}

\begin{figure*}
\includegraphics[width=0.45\textwidth]{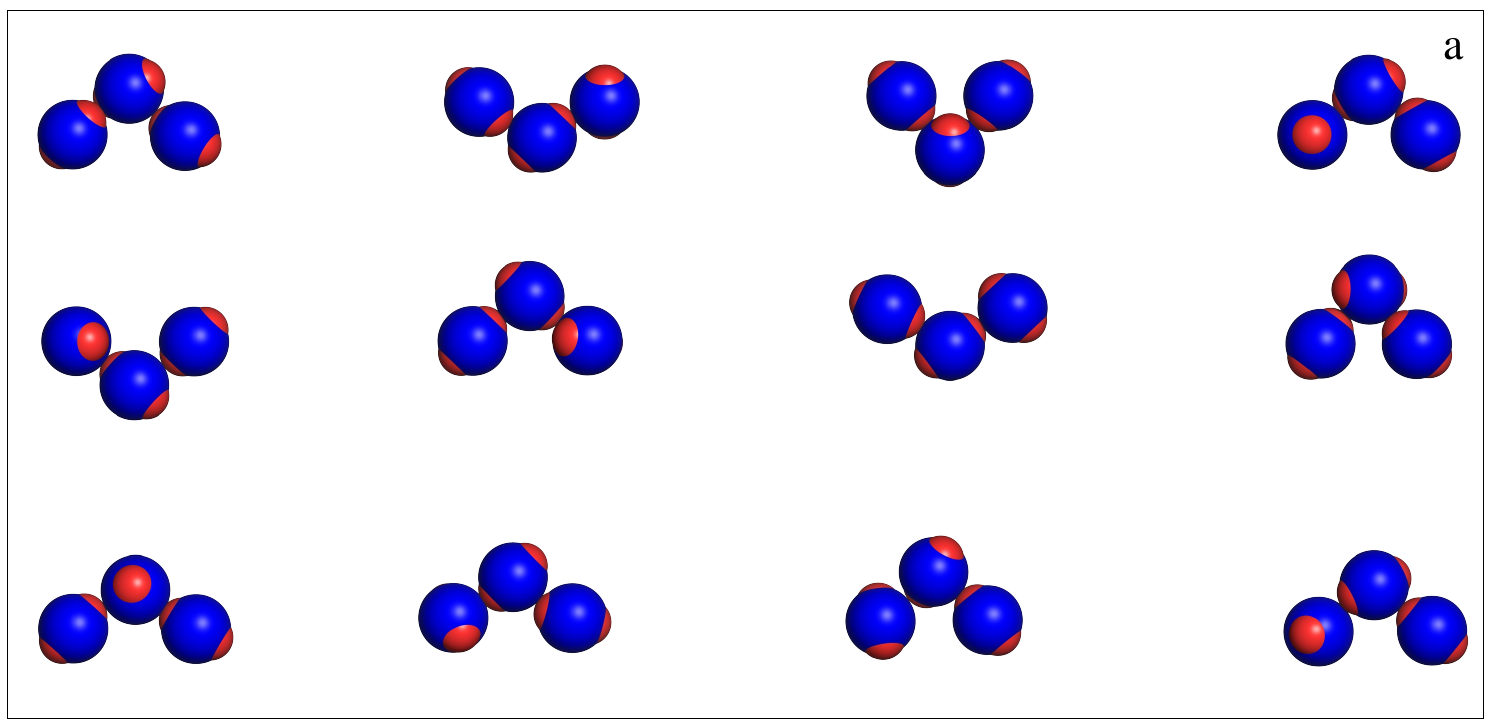}
\includegraphics[width=0.45\textwidth]{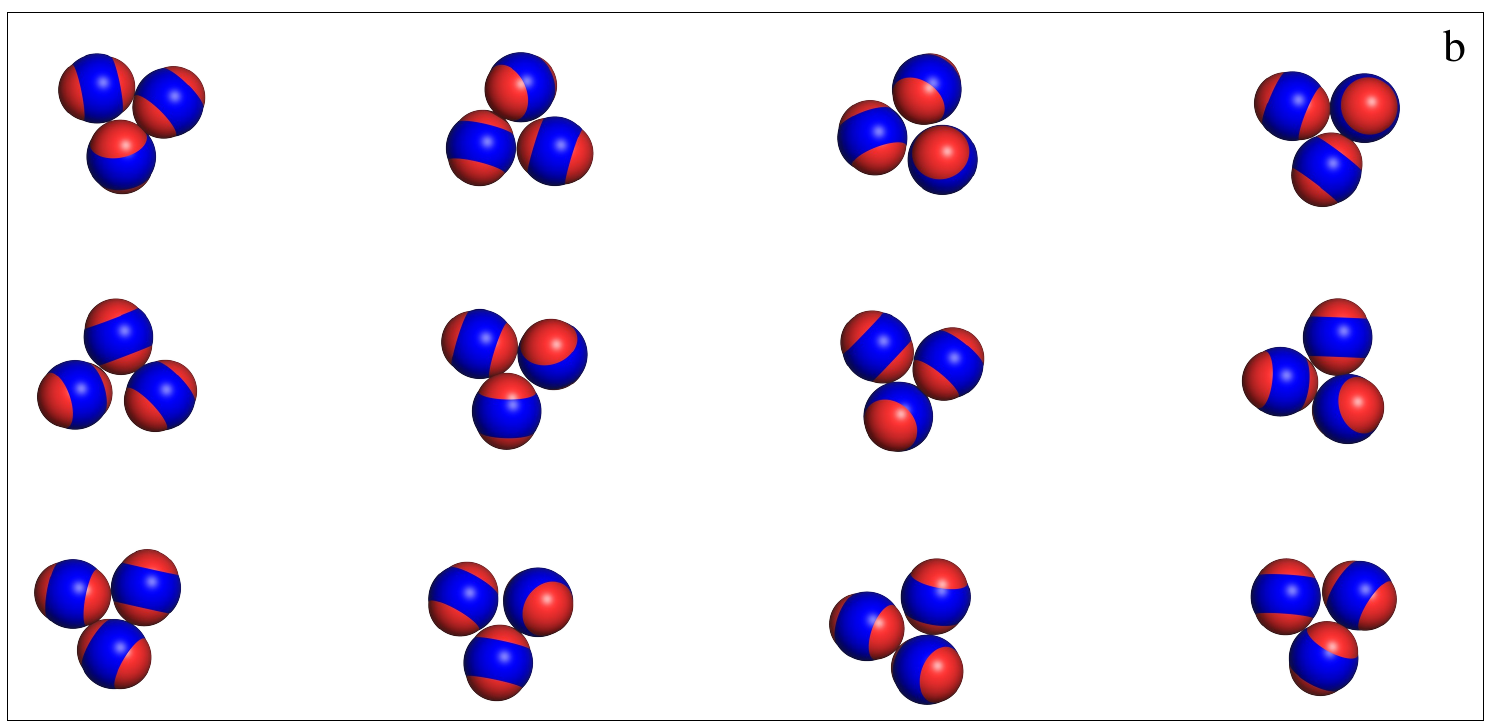}
\caption{Compact vs branched trimers at criticality for $u_{\rm EE}=u_{\rm PP}=0.0$ ($\rm IPP_{\rm ro}$). (a) Typical configurations of branched trimers for $\gamma=30$°. Samples from simulations at the inferred critical point. (b) Typical configurations of compact trimers for $\gamma=55$°. Samples from simulations
at the inferred critical point.
}
\label{fig:trimers}
\end{figure*}

Fig.~\ref{fig:xi_k} displays $\xi_0$, $\xi_1$ and
$\xi_2$ for $u_{\rm EE} = u_{\rm PP}=0.0$ and $u_{\rm EE} = 0.5$, $u_{\rm PP}=2.0$. The summation in Eq.~(\ref{eq:xi_k})
has been bounded to a maximal value of the cluster size as the number of extremely large clusters is not statistically significant,
but these clusters would nonetheless dominate the summation, making the measure of $\xi$ too noisy. For both the systems considered $\xi_k$ is a monotonically decreasing function of $\gamma$, regardless of $k$. Furthermore, data for the repulsive system $u_{\rm EE} = 0.5$, $u_{\rm PP}=2.0$ ($\rm IPP_{\rm ref}$) tend to display larger 
values of $\xi_k$ if compared with the data for the non-repulsive system 
$u_{\rm EE} = u_{\rm PP}=0.0$ ($\rm IPP_{\rm ro}$) at the same $\gamma$. These results are perfectly consistent with the behaviour observed for the critical density $\rho_c$: models characterized by large $\rho_c$ tend to have large $\xi_k$ and vice versa, confirming the interpretation given in the main text, i.e., that  the behaviour of the density can be explained in terms of the compactness of the clusters. Compact cluster have indeed a smaller radius of gyration and the measure of $\xi$ confirms that $\rho_c$ is large for systems where  compact structures are common.
Fig.~\ref{fig:trimers} shows  examples of branched trimers for $\gamma=30$° and compact trimers for $\gamma=55$°, 
clarifying why the gyration radius of clusters with large $\gamma$ tend to be
smaller.

\section{Calculation of the bonding volume}

The bonding volume is defined as
\begin{equation}
    V_b = 4\pi \int_{2\sigma_c}^{2\sigma_c + \delta} S(r) r^2 dr 
\end{equation}
where $S(r)$ is the average fraction of solid angle available to bonding~\cite{Sciortino2007Self}.
To compute $S(r)$, a pair of particles was simulated, one being fixed in position and orientation, while the other being assigned a random orientation and a
random position in the shell of radii $2 \sigma_c$ and $2 \sigma_c + \delta$.
This range of distances ensures the existence of a random geometric bond. $S(r)$ is then computed as
the fraction of configurations in which the geometric bond is also an energetic
bond, averaged at different values of $r$.

\begin{figure*}
\includegraphics[width=0.92\textwidth]{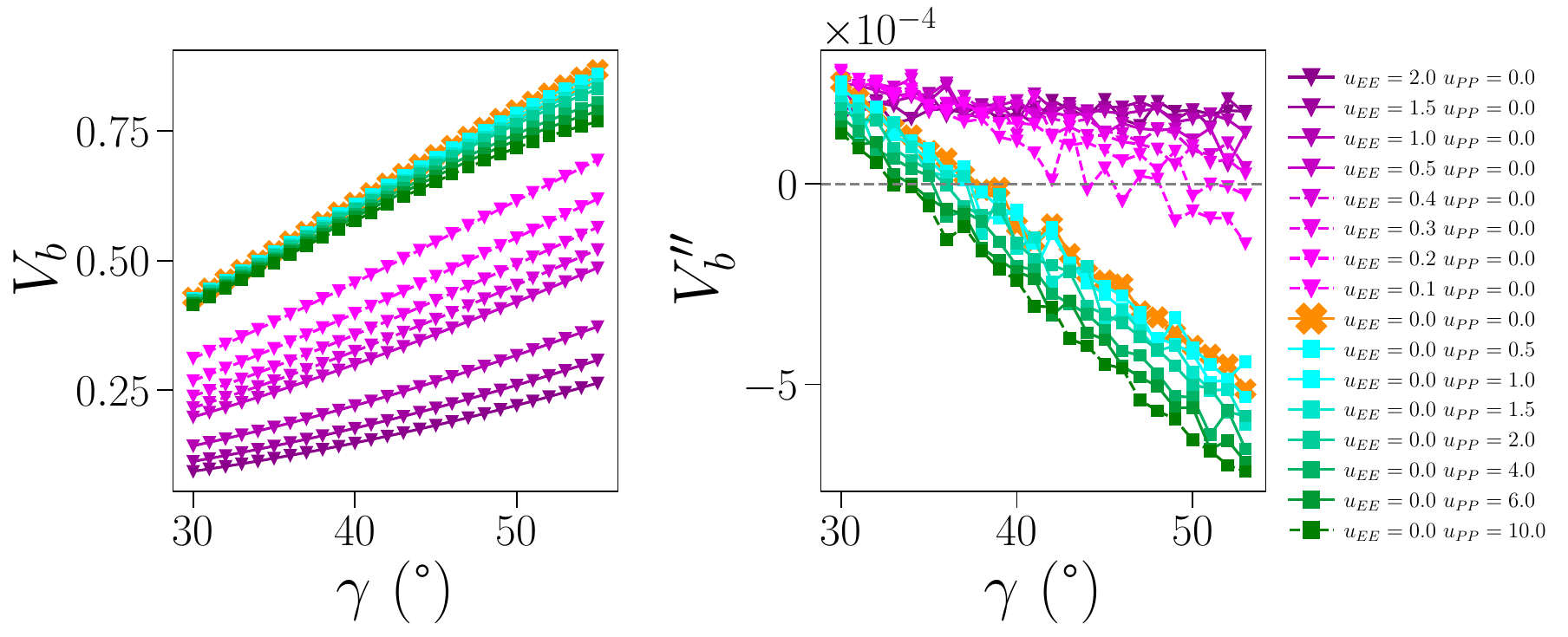}
\caption{Bonding volume $V_b$ and its second derivative w.r.t. $\gamma$ for
different interaction strengths. Orange crosses are the data corresponding to 
$u_{\rm EE} = u_{\rm PP}=0$ ($\rm IPP_{\rm ro}$).  The color goes from magenta to purple as $u_{\rm EE}$ grows
and from cyan to green as $u_{\rm PP}$ grows. Curves referring to values of $u_{\rm EE}$ and $u_{\rm PP}$ for which we computed the critical point are shown as solid lines, otherwise the line is dashed.}
\label{fig:Vb}
\end{figure*}

In Fig.~\ref{fig:Vb} the bonding volume is shown, together with its second derivative w.r.t. $\gamma$. 

Consistently with the behaviour of the critical temperature reported in the main text,  $V_b$ is found to monotonically increase with $\gamma$. Furthermore, $V_b$ decreases as electrostatic repulsion grows, with $u_{\rm EE}$ having more effect than $u_{\rm PP}$. The curvature of $V_b$ versus $\gamma$, expressed by the second derivative, decreases with $\gamma$ -- quickly with $u_{\rm PP}$ changes, slowly $u_{\rm EE}$ changes -- and it can change sign the $\gamma$. This is the same trend shown by the critical temperature, see Fig.2d of the main text. For $\gamma=30$ the dependence of $V_b$ on $u_{\rm PP}$ is rather weak, as it is for $T_c$ but, curiously, $V_b$ remains a (slighlty) decreasing function, while $T_c$ slowly increases with $u_{\rm PP}$.

% \bibliography{biblio}

\begin{thebibliography}{78}%
\makeatletter
\providecommand \@ifxundefined [1]{%
 \@ifx{#1\undefined}
}%
\providecommand \@ifnum [1]{%
 \ifnum #1\expandafter \@firstoftwo
 \else \expandafter \@secondoftwo
 \fi
}%
\providecommand \@ifx [1]{%
 \ifx #1\expandafter \@firstoftwo
 \else \expandafter \@secondoftwo
 \fi
}%
\providecommand \natexlab [1]{#1}%
\providecommand \enquote  [1]{``#1''}%
\providecommand \bibnamefont  [1]{#1}%
\providecommand \bibfnamefont [1]{#1}%
\providecommand \citenamefont [1]{#1}%
\providecommand \href@noop [0]{\@secondoftwo}%
\providecommand \href [0]{\begingroup \@sanitize@url \@href}%
\providecommand \@href[1]{\@@startlink{#1}\@@href}%
\providecommand \@@href[1]{\endgroup#1\@@endlink}%
\providecommand \@sanitize@url [0]{\catcode `\\12\catcode `\$12\catcode `\&12\catcode `\#12\catcode `\^12\catcode `\_12\catcode `\%12\relax}%
\providecommand \@@startlink[1]{}%
\providecommand \@@endlink[0]{}%
\providecommand \url  [0]{\begingroup\@sanitize@url \@url }%
\providecommand \@url [1]{\endgroup\@href {#1}{\urlprefix }}%
\providecommand \urlprefix  [0]{URL }%
\providecommand \Eprint [0]{\href }%
\providecommand \doibase [0]{https://doi.org/}%
\providecommand \selectlanguage [0]{\@gobble}%
\providecommand \bibinfo  [0]{\@secondoftwo}%
\providecommand \bibfield  [0]{\@secondoftwo}%
\providecommand \translation [1]{[#1]}%
\providecommand \BibitemOpen [0]{}%
\providecommand \bibitemStop [0]{}%
\providecommand \bibitemNoStop [0]{.\EOS\space}%
\providecommand \EOS [0]{\spacefactor3000\relax}%
\providecommand \BibitemShut  [1]{\csname bibitem#1\endcsname}%
\let\auto@bib@innerbib\@empty
%</preamble>
\bibitem [{\citenamefont {Gunton}\ \emph {et~al.}(2007)\citenamefont {Gunton}, \citenamefont {Shiryayev},\ and\ \citenamefont {Pagan}}]{protein_book_2007}%
  \BibitemOpen
  \bibfield  {author} {\bibinfo {author} {\bibfnamefont {J.~D.}\ \bibnamefont {Gunton}}, \bibinfo {author} {\bibfnamefont {A.}~\bibnamefont {Shiryayev}},\ and\ \bibinfo {author} {\bibfnamefont {D.~L.}\ \bibnamefont {Pagan}},\ }\href@noop {} {\emph {\bibinfo {title} {Protein Condensation: Kinetic Pathways to Crystallization and Disease}}}\ (\bibinfo  {publisher} {Cambridge University Press},\ \bibinfo {year} {2007})\BibitemShut {NoStop}%
\bibitem [{\citenamefont {McManus}\ \emph {et~al.}(2016)\citenamefont {McManus}, \citenamefont {Charbonneau}, \citenamefont {Zaccarelli},\ and\ \citenamefont {Asherie}}]{McManus2016The}%
  \BibitemOpen
  \bibfield  {author} {\bibinfo {author} {\bibfnamefont {J.~J.}\ \bibnamefont {McManus}}, \bibinfo {author} {\bibfnamefont {P.}~\bibnamefont {Charbonneau}}, \bibinfo {author} {\bibfnamefont {E.}~\bibnamefont {Zaccarelli}},\ and\ \bibinfo {author} {\bibfnamefont {N.}~\bibnamefont {Asherie}},\ }\bibfield  {title} {\bibinfo {title} {The physics of protein self-assembly},\ }\href@noop {} {\bibfield  {journal} {\bibinfo  {journal} {Current Opinion in Colloid \& Interface Science}\ }\textbf {\bibinfo {volume} {22}},\ \bibinfo {pages} {73} (\bibinfo {year} {2016})}\BibitemShut {NoStop}%
\bibitem [{\citenamefont {Lomakin}\ \emph {et~al.}(1999)\citenamefont {Lomakin}, \citenamefont {Asherie},\ and\ \citenamefont {Benedek}}]{Benedek_1999}%
  \BibitemOpen
  \bibfield  {author} {\bibinfo {author} {\bibfnamefont {A.}~\bibnamefont {Lomakin}}, \bibinfo {author} {\bibfnamefont {N.}~\bibnamefont {Asherie}},\ and\ \bibinfo {author} {\bibfnamefont {G.~B.}\ \bibnamefont {Benedek}},\ }\bibfield  {title} {\bibinfo {title} {Aeolotopic interactions of globular proteins},\ }\href@noop {} {\bibfield  {journal} {\bibinfo  {journal} {Proceedings of the National Academy of Sciences}\ }\textbf {\bibinfo {volume} {96}},\ \bibinfo {pages} {9465} (\bibinfo {year} {1999})}\BibitemShut {NoStop}%
\bibitem [{\citenamefont {Gibaud}\ \emph {et~al.}(2011)\citenamefont {Gibaud}, \citenamefont {Cardinaux}, \citenamefont {Bergenholtz}, \citenamefont {Stradner},\ and\ \citenamefont {Schurtenberger}}]{Schurtenberger_2011}%
  \BibitemOpen
  \bibfield  {author} {\bibinfo {author} {\bibfnamefont {T.}~\bibnamefont {Gibaud}}, \bibinfo {author} {\bibfnamefont {F.}~\bibnamefont {Cardinaux}}, \bibinfo {author} {\bibfnamefont {J.}~\bibnamefont {Bergenholtz}}, \bibinfo {author} {\bibfnamefont {A.}~\bibnamefont {Stradner}},\ and\ \bibinfo {author} {\bibfnamefont {P.}~\bibnamefont {Schurtenberger}},\ }\bibfield  {title} {\bibinfo {title} {Phase separation and dynamical arrest for particles interacting with mixed potentials—the case of globular proteins revisited},\ }\href@noop {} {\bibfield  {journal} {\bibinfo  {journal} {Soft Matter}\ }\textbf {\bibinfo {volume} {7}},\ \bibinfo {pages} {857} (\bibinfo {year} {2011})}\BibitemShut {NoStop}%
\bibitem [{\citenamefont {Platten}\ \emph {et~al.}(2015)\citenamefont {Platten}, \citenamefont {Valadez-P{\'e}rez}, \citenamefont {Casta{\~n}eda-Priego},\ and\ \citenamefont {Egelhaaf}}]{Egelhaaf_2015}%
  \BibitemOpen
  \bibfield  {author} {\bibinfo {author} {\bibfnamefont {F.}~\bibnamefont {Platten}}, \bibinfo {author} {\bibfnamefont {N.~E.}\ \bibnamefont {Valadez-P{\'e}rez}}, \bibinfo {author} {\bibfnamefont {R.}~\bibnamefont {Casta{\~n}eda-Priego}},\ and\ \bibinfo {author} {\bibfnamefont {S.~U.}\ \bibnamefont {Egelhaaf}},\ }\bibfield  {title} {\bibinfo {title} {Extended law of corresponding states for protein solutions},\ }\href@noop {} {\bibfield  {journal} {\bibinfo  {journal} {J. Chem. Phys.}\ }\textbf {\bibinfo {volume} {142}},\ \bibinfo {pages} {174905} (\bibinfo {year} {2015})}\BibitemShut {NoStop}%
\bibitem [{\citenamefont {Hagen}\ and\ \citenamefont {Frenkel}(1994)}]{HagenFrenkel_1994}%
  \BibitemOpen
  \bibfield  {author} {\bibinfo {author} {\bibfnamefont {M.~H.~J.}\ \bibnamefont {Hagen}}\ and\ \bibinfo {author} {\bibfnamefont {D.}~\bibnamefont {Frenkel}},\ }\bibfield  {title} {\bibinfo {title} {Determination of phase diagrams for the hard‐core attractive {Yukawa} system},\ }\href@noop {} {\bibfield  {journal} {\bibinfo  {journal} {J. Chem. Phys.}\ }\textbf {\bibinfo {volume} {101}},\ \bibinfo {pages} {4093} (\bibinfo {year} {1994})}\BibitemShut {NoStop}%
\bibitem [{\citenamefont {Miller}\ and\ \citenamefont {Frenkel}(2004)}]{MillerFrenkel_2004}%
  \BibitemOpen
  \bibfield  {author} {\bibinfo {author} {\bibfnamefont {M.~A.}\ \bibnamefont {Miller}}\ and\ \bibinfo {author} {\bibfnamefont {D.}~\bibnamefont {Frenkel}},\ }\bibfield  {title} {\bibinfo {title} {Phase diagram of the adhesive hard sphere fluid},\ }\href@noop {} {\bibfield  {journal} {\bibinfo  {journal} {J. Chem. Phys.}\ }\textbf {\bibinfo {volume} {121}},\ \bibinfo {pages} {535} (\bibinfo {year} {2004})}\BibitemShut {NoStop}%
\bibitem [{\citenamefont {Pagan}\ and\ \citenamefont {Gunton}(2005)}]{PaganGunton_2005}%
  \BibitemOpen
  \bibfield  {author} {\bibinfo {author} {\bibfnamefont {D.~L.}\ \bibnamefont {Pagan}}\ and\ \bibinfo {author} {\bibfnamefont {J.~D.}\ \bibnamefont {Gunton}},\ }\bibfield  {title} {\bibinfo {title} {Phase behavior of short-range square-well model},\ }\href@noop {} {\bibfield  {journal} {\bibinfo  {journal} {J. Chem. Phys.}\ }\textbf {\bibinfo {volume} {122}},\ \bibinfo {pages} {184515} (\bibinfo {year} {2005})}\BibitemShut {NoStop}%
\bibitem [{\citenamefont {{ten Wolde}}\ and\ \citenamefont {Frenkel}(1997)}]{tenWoldeFrenkel_1997}%
  \BibitemOpen
  \bibfield  {author} {\bibinfo {author} {\bibfnamefont {P.~R.}\ \bibnamefont {{ten Wolde}}}\ and\ \bibinfo {author} {\bibfnamefont {D.}~\bibnamefont {Frenkel}},\ }\bibfield  {title} {\bibinfo {title} {Enhancement of protein crystal nucleation by critical density fluctuations},\ }\href@noop {} {\bibfield  {journal} {\bibinfo  {journal} {Science}\ }\textbf {\bibinfo {volume} {277}},\ \bibinfo {pages} {1975} (\bibinfo {year} {1997})}\BibitemShut {NoStop}%
\bibitem [{\citenamefont {Vliegenthart}\ and\ \citenamefont {Lekkerkerker}(2000)}]{VliegenthartLekkerkerker_2000}%
  \BibitemOpen
  \bibfield  {author} {\bibinfo {author} {\bibfnamefont {G.~A.}\ \bibnamefont {Vliegenthart}}\ and\ \bibinfo {author} {\bibfnamefont {H.~N.~W.}\ \bibnamefont {Lekkerkerker}},\ }\bibfield  {title} {\bibinfo {title} {Predicting the gas–liquid critical point from the second virial coefficient},\ }\href@noop {} {\bibfield  {journal} {\bibinfo  {journal} {J. Chem. Phys}\ }\textbf {\bibinfo {volume} {112}},\ \bibinfo {pages} {5364} (\bibinfo {year} {2000})}\BibitemShut {NoStop}%
\bibitem [{\citenamefont {de~Hoog}\ \emph {et~al.}(2001)\citenamefont {de~Hoog}, \citenamefont {Kegel}, \citenamefont {van Blaaderen},\ and\ \citenamefont {Lekkerkerker}}]{deHoog_2001}%
  \BibitemOpen
  \bibfield  {author} {\bibinfo {author} {\bibfnamefont {E.~H.~A.}\ \bibnamefont {de~Hoog}}, \bibinfo {author} {\bibfnamefont {W.~K.}\ \bibnamefont {Kegel}}, \bibinfo {author} {\bibfnamefont {A.}~\bibnamefont {van Blaaderen}},\ and\ \bibinfo {author} {\bibfnamefont {H.~N.~W.}\ \bibnamefont {Lekkerkerker}},\ }\bibfield  {title} {\bibinfo {title} {Direct observation of crystallization and aggregation in a phase-separating colloid-polymer suspension},\ }\href@noop {} {\bibfield  {journal} {\bibinfo  {journal} {Phys. Rev. E}\ }\textbf {\bibinfo {volume} {64}},\ \bibinfo {pages} {021407} (\bibinfo {year} {2001})}\BibitemShut {NoStop}%
\bibitem [{\citenamefont {Anderson}\ and\ \citenamefont {Lekkerkerker}(2002)}]{AndersonLekkerkerker_2002}%
  \BibitemOpen
  \bibfield  {author} {\bibinfo {author} {\bibfnamefont {V.~J.}\ \bibnamefont {Anderson}}\ and\ \bibinfo {author} {\bibfnamefont {H.~N.~W.}\ \bibnamefont {Lekkerkerker}},\ }\bibfield  {title} {\bibinfo {title} {Insights into phase transition kinetics from colloid science},\ }\href@noop {} {\bibfield  {journal} {\bibinfo  {journal} {Nature}\ }\textbf {\bibinfo {volume} {416}},\ \bibinfo {pages} {811} (\bibinfo {year} {2002})}\BibitemShut {NoStop}%
\bibitem [{\citenamefont {Sear}(1999)}]{Sear_1999}%
  \BibitemOpen
  \bibfield  {author} {\bibinfo {author} {\bibfnamefont {R.~P.}\ \bibnamefont {Sear}},\ }\bibfield  {title} {\bibinfo {title} {Phase behavior of a simple model of globular proteins},\ }\href@noop {} {\bibfield  {journal} {\bibinfo  {journal} {J. Chem. Phys.}\ }\textbf {\bibinfo {volume} {111}},\ \bibinfo {pages} {4800} (\bibinfo {year} {1999})}\BibitemShut {NoStop}%
\bibitem [{\citenamefont {Kern}\ and\ \citenamefont {Frenkel}(2003)}]{KernFrenkel_2003}%
  \BibitemOpen
  \bibfield  {author} {\bibinfo {author} {\bibfnamefont {N.}~\bibnamefont {Kern}}\ and\ \bibinfo {author} {\bibfnamefont {D.}~\bibnamefont {Frenkel}},\ }\bibfield  {title} {\bibinfo {title} {Fluid–fluid coexistence in colloidal systems with short-ranged strongly directional attraction},\ }\href@noop {} {\bibfield  {journal} {\bibinfo  {journal} {J. Chem. Phys.}\ }\textbf {\bibinfo {volume} {118}},\ \bibinfo {pages} {9882} (\bibinfo {year} {2003})}\BibitemShut {NoStop}%
\bibitem [{\citenamefont {Bianchi}\ \emph {et~al.}(2006)\citenamefont {Bianchi}, \citenamefont {Largo}, \citenamefont {Tartaglia}, \citenamefont {Zaccarelli},\ and\ \citenamefont {Sciortino}}]{Bianchi2006Phase}%
  \BibitemOpen
  \bibfield  {author} {\bibinfo {author} {\bibfnamefont {E.}~\bibnamefont {Bianchi}}, \bibinfo {author} {\bibfnamefont {J.}~\bibnamefont {Largo}}, \bibinfo {author} {\bibfnamefont {P.}~\bibnamefont {Tartaglia}}, \bibinfo {author} {\bibfnamefont {E.}~\bibnamefont {Zaccarelli}},\ and\ \bibinfo {author} {\bibfnamefont {F.}~\bibnamefont {Sciortino}},\ }\bibfield  {title} {\bibinfo {title} {Phase diagram of patchy colloids: Towards empty liquids},\ }\href {https://doi.org/10.1103/PhysRevLett.97.168301} {\bibfield  {journal} {\bibinfo  {journal} {Phys. Rev. Lett.}\ }\textbf {\bibinfo {volume} {97}},\ \bibinfo {pages} {168301} (\bibinfo {year} {2006})}\BibitemShut {NoStop}%
\bibitem [{\citenamefont {Espinosa}\ \emph {et~al.}(2020)\citenamefont {Espinosa}, \citenamefont {Joseph}, \citenamefont {Sanchez-Burgos}, \citenamefont {Garaizar}, \citenamefont {Frenkel},\ and\ \citenamefont {Collepardo-Guevara}}]{Espinoza2020Liquid}%
  \BibitemOpen
  \bibfield  {author} {\bibinfo {author} {\bibfnamefont {J.~R.}\ \bibnamefont {Espinosa}}, \bibinfo {author} {\bibfnamefont {J.~A.}\ \bibnamefont {Joseph}}, \bibinfo {author} {\bibfnamefont {I.}~\bibnamefont {Sanchez-Burgos}}, \bibinfo {author} {\bibfnamefont {A.}~\bibnamefont {Garaizar}}, \bibinfo {author} {\bibfnamefont {D.}~\bibnamefont {Frenkel}},\ and\ \bibinfo {author} {\bibfnamefont {R.}~\bibnamefont {Collepardo-Guevara}},\ }\bibfield  {title} {\bibinfo {title} {Liquid network connectivity regulates the stability and composition of biomolecular condensates with many components},\ }\href@noop {} {\bibfield  {journal} {\bibinfo  {journal} {Proceedings of the National Academy of Sciences}\ }\textbf {\bibinfo {volume} {117}},\ \bibinfo {pages} {13238} (\bibinfo {year} {2020})}\BibitemShut {NoStop}%
\bibitem [{\citenamefont {Roosen-Runge}\ \emph {et~al.}(2014)\citenamefont {Roosen-Runge}, \citenamefont {Zhang}, \citenamefont {Schreiber},\ and\ \citenamefont {Roth}}]{roosenrunge2014sr}%
  \BibitemOpen
  \bibfield  {author} {\bibinfo {author} {\bibfnamefont {F.}~\bibnamefont {Roosen-Runge}}, \bibinfo {author} {\bibfnamefont {F.}~\bibnamefont {Zhang}}, \bibinfo {author} {\bibfnamefont {F.}~\bibnamefont {Schreiber}},\ and\ \bibinfo {author} {\bibfnamefont {R.}~\bibnamefont {Roth}},\ }\bibfield  {title} {\bibinfo {title} {Ion-activated attractive patches as a mechanism for controlled protein interactions},\ }\href {https://doi.org/10.1038/srep07016} {\bibfield  {journal} {\bibinfo  {journal} {Scientific Reports}\ }\textbf {\bibinfo {volume} {4}},\ \bibinfo {pages} {7016} (\bibinfo {year} {2014})}\BibitemShut {NoStop}%
\bibitem [{\citenamefont {Cai}\ \emph {et~al.}(2017)\citenamefont {Cai}, \citenamefont {Townsend}, \citenamefont {Dodson}, \citenamefont {Heiney},\ and\ \citenamefont {Sweeney}}]{Sweeney_2017}%
  \BibitemOpen
  \bibfield  {author} {\bibinfo {author} {\bibfnamefont {J.}~\bibnamefont {Cai}}, \bibinfo {author} {\bibfnamefont {J.~P.}\ \bibnamefont {Townsend}}, \bibinfo {author} {\bibfnamefont {T.~C.}\ \bibnamefont {Dodson}}, \bibinfo {author} {\bibfnamefont {P.~A.}\ \bibnamefont {Heiney}},\ and\ \bibinfo {author} {\bibfnamefont {A.~M.}\ \bibnamefont {Sweeney}},\ }\bibfield  {title} {\bibinfo {title} {Eye patches: Protein assembly of index-gradient squid lenses},\ }\href@noop {} {\bibfield  {journal} {\bibinfo  {journal} {Science}\ }\textbf {\bibinfo {volume} {357}},\ \bibinfo {pages} {564} (\bibinfo {year} {2017})}\BibitemShut {NoStop}%
\bibitem [{\citenamefont {Heidenreich}\ \emph {et~al.}(2020)\citenamefont {Heidenreich}, \citenamefont {Georgeson}, \citenamefont {Locatelli}, \citenamefont {Rovigatti}, \citenamefont {Nandi}, \citenamefont {Steinberg}, \citenamefont {Nadav}, \citenamefont {Shimoni}, \citenamefont {Safran}, \citenamefont {Doye},\ and\ \citenamefont {Levy}}]{Levy_2020}%
  \BibitemOpen
  \bibfield  {author} {\bibinfo {author} {\bibfnamefont {M.}~\bibnamefont {Heidenreich}}, \bibinfo {author} {\bibfnamefont {J.~M.}\ \bibnamefont {Georgeson}}, \bibinfo {author} {\bibfnamefont {E.}~\bibnamefont {Locatelli}}, \bibinfo {author} {\bibfnamefont {L.}~\bibnamefont {Rovigatti}}, \bibinfo {author} {\bibfnamefont {S.~K.}\ \bibnamefont {Nandi}}, \bibinfo {author} {\bibfnamefont {A.}~\bibnamefont {Steinberg}}, \bibinfo {author} {\bibfnamefont {Y.}~\bibnamefont {Nadav}}, \bibinfo {author} {\bibfnamefont {E.}~\bibnamefont {Shimoni}}, \bibinfo {author} {\bibfnamefont {S.~A.}\ \bibnamefont {Safran}}, \bibinfo {author} {\bibfnamefont {J.~P.~K.}\ \bibnamefont {Doye}},\ and\ \bibinfo {author} {\bibfnamefont {E.~D.}\ \bibnamefont {Levy}},\ }\bibfield  {title} {\bibinfo {title} {Designer protein assemblies with tunable phase diagrams in living cells},\ }\href@noop {} {\bibfield  {journal} {\bibinfo  {journal} {Nature Chemical Biology}\ }\textbf {\bibinfo {volume} {16}},\ \bibinfo {pages} {939} (\bibinfo {year}
  {2020})}\BibitemShut {NoStop}%
\bibitem [{\citenamefont {Zhang}\ and\ \citenamefont {Glotzer}(2004)}]{Glotzer_2004}%
  \BibitemOpen
  \bibfield  {author} {\bibinfo {author} {\bibfnamefont {Z.}~\bibnamefont {Zhang}}\ and\ \bibinfo {author} {\bibfnamefont {S.~C.}\ \bibnamefont {Glotzer}},\ }\bibfield  {title} {\bibinfo {title} {Self-assembly of patchy particles},\ }\href@noop {} {\bibfield  {journal} {\bibinfo  {journal} {Nano Lett.}\ }\textbf {\bibinfo {volume} {4}},\ \bibinfo {pages} {1407} (\bibinfo {year} {2004})}\BibitemShut {NoStop}%
\bibitem [{\citenamefont {Pawar}\ and\ \citenamefont {Kretzschmar}(2010)}]{Kretzschmar_2010}%
  \BibitemOpen
  \bibfield  {author} {\bibinfo {author} {\bibfnamefont {A.}~\bibnamefont {Pawar}}\ and\ \bibinfo {author} {\bibfnamefont {I.}~\bibnamefont {Kretzschmar}},\ }\bibfield  {title} {\bibinfo {title} {Fabrication, assembly, and application of patchy particles},\ }\href@noop {} {\bibfield  {journal} {\bibinfo  {journal} {Macromol. Rapid. Commun.}\ }\textbf {\bibinfo {volume} {31}},\ \bibinfo {pages} {150} (\bibinfo {year} {2010})}\BibitemShut {NoStop}%
\bibitem [{\citenamefont {Bianchi}\ \emph {et~al.}(2011{\natexlab{a}})\citenamefont {Bianchi}, \citenamefont {Blaak},\ and\ \citenamefont {Likos}}]{Bianchi_pccp_2011}%
  \BibitemOpen
  \bibfield  {author} {\bibinfo {author} {\bibfnamefont {E.}~\bibnamefont {Bianchi}}, \bibinfo {author} {\bibfnamefont {R.}~\bibnamefont {Blaak}},\ and\ \bibinfo {author} {\bibfnamefont {C.~N.}\ \bibnamefont {Likos}},\ }\bibfield  {title} {\bibinfo {title} {Patchy colloids: state of the art and perspectives},\ }\href@noop {} {\bibfield  {journal} {\bibinfo  {journal} {Phys. Chem. Chem. Phys.}\ }\textbf {\bibinfo {volume} {13}},\ \bibinfo {pages} {6397} (\bibinfo {year} {2011}{\natexlab{a}})}\BibitemShut {NoStop}%
\bibitem [{\citenamefont {Wang}\ \emph {et~al.}(2012)\citenamefont {Wang}, \citenamefont {Wang}, \citenamefont {Breed}, \citenamefont {Manoharan}, \citenamefont {Feng}, \citenamefont {Hollingsworth}, \citenamefont {Weck},\ and\ \citenamefont {Pine}}]{Pine_2012}%
  \BibitemOpen
  \bibfield  {author} {\bibinfo {author} {\bibfnamefont {Y.}~\bibnamefont {Wang}}, \bibinfo {author} {\bibfnamefont {Y.}~\bibnamefont {Wang}}, \bibinfo {author} {\bibfnamefont {D.~R.}\ \bibnamefont {Breed}}, \bibinfo {author} {\bibfnamefont {V.~N.}\ \bibnamefont {Manoharan}}, \bibinfo {author} {\bibfnamefont {L.}~\bibnamefont {Feng}}, \bibinfo {author} {\bibfnamefont {A.~D.}\ \bibnamefont {Hollingsworth}}, \bibinfo {author} {\bibfnamefont {M.}~\bibnamefont {Weck}},\ and\ \bibinfo {author} {\bibfnamefont {D.~J.}\ \bibnamefont {Pine}},\ }\bibfield  {title} {\bibinfo {title} {Colloids with valence and specific directional bonding},\ }\href@noop {} {\bibfield  {journal} {\bibinfo  {journal} {Nature}\ }\textbf {\bibinfo {volume} {491}},\ \bibinfo {pages} {51} (\bibinfo {year} {2012})}\BibitemShut {NoStop}%
\bibitem [{\citenamefont {Gong}\ \emph {et~al.}(2017)\citenamefont {Gong}, \citenamefont {Hueckel}, \citenamefont {Yi},\ and\ \citenamefont {Sacanna}}]{Sacanna_2017}%
  \BibitemOpen
  \bibfield  {author} {\bibinfo {author} {\bibfnamefont {Z.}~\bibnamefont {Gong}}, \bibinfo {author} {\bibfnamefont {T.}~\bibnamefont {Hueckel}}, \bibinfo {author} {\bibfnamefont {G.-R.}\ \bibnamefont {Yi}},\ and\ \bibinfo {author} {\bibfnamefont {S.}~\bibnamefont {Sacanna}},\ }\bibfield  {title} {\bibinfo {title} {Patchy particles made by colloidal fusion},\ }\href@noop {} {\bibfield  {journal} {\bibinfo  {journal} {Nature}\ }\textbf {\bibinfo {volume} {550}},\ \bibinfo {pages} {234} (\bibinfo {year} {2017})}\BibitemShut {NoStop}%
\bibitem [{\citenamefont {Chen}\ \emph {et~al.}(2011)\citenamefont {Chen}, \citenamefont {Diesel}, \citenamefont {Whitmer}, \citenamefont {Bae}, \citenamefont {Luijten},\ and\ \citenamefont {Granick}}]{Granick_2011}%
  \BibitemOpen
  \bibfield  {author} {\bibinfo {author} {\bibfnamefont {Q.}~\bibnamefont {Chen}}, \bibinfo {author} {\bibfnamefont {E.}~\bibnamefont {Diesel}}, \bibinfo {author} {\bibfnamefont {J.~K.}\ \bibnamefont {Whitmer}}, \bibinfo {author} {\bibfnamefont {S.~C.}\ \bibnamefont {Bae}}, \bibinfo {author} {\bibfnamefont {E.}~\bibnamefont {Luijten}},\ and\ \bibinfo {author} {\bibfnamefont {S.}~\bibnamefont {Granick}},\ }\bibfield  {title} {\bibinfo {title} {Triblock colloids for directed self-assembly},\ }\href@noop {} {\bibfield  {journal} {\bibinfo  {journal} {J. Am. Chem. Soc.}\ }\textbf {\bibinfo {volume} {133}},\ \bibinfo {pages} {7725} (\bibinfo {year} {2011})}\BibitemShut {NoStop}%
\bibitem [{\citenamefont {Bianchi}\ \emph {et~al.}(2017)\citenamefont {Bianchi}, \citenamefont {Capone}, \citenamefont {Coluzza}, \citenamefont {Rovigatti},\ and\ \citenamefont {van Oostrum}}]{Bianchi2017Limiting}%
  \BibitemOpen
  \bibfield  {author} {\bibinfo {author} {\bibfnamefont {E.}~\bibnamefont {Bianchi}}, \bibinfo {author} {\bibfnamefont {B.}~\bibnamefont {Capone}}, \bibinfo {author} {\bibfnamefont {I.}~\bibnamefont {Coluzza}}, \bibinfo {author} {\bibfnamefont {L.}~\bibnamefont {Rovigatti}},\ and\ \bibinfo {author} {\bibfnamefont {P.~D.~J.}\ \bibnamefont {van Oostrum}},\ }\bibfield  {title} {\bibinfo {title} {Limiting the valence: advancements and new perspectives on patchy colloids{,} soft functionalized nanoparticles and biomolecules},\ }\href {https://doi.org/10.1039/C7CP03149A} {\bibfield  {journal} {\bibinfo  {journal} {Phys. Chem. Chem. Phys.}\ }\textbf {\bibinfo {volume} {19}},\ \bibinfo {pages} {19847} (\bibinfo {year} {2017})}\BibitemShut {NoStop}%
\bibitem [{\citenamefont {He}\ \emph {et~al.}(2020)\citenamefont {He}, \citenamefont {Gales}, \citenamefont {Ducrot}, \citenamefont {Gong}, \citenamefont {Yi}, \citenamefont {Sacanna},\ and\ \citenamefont {Pine}}]{Pine_2020}%
  \BibitemOpen
  \bibfield  {author} {\bibinfo {author} {\bibfnamefont {M.}~\bibnamefont {He}}, \bibinfo {author} {\bibfnamefont {J.~P.}\ \bibnamefont {Gales}}, \bibinfo {author} {\bibfnamefont {{\'E}.}~\bibnamefont {Ducrot}}, \bibinfo {author} {\bibfnamefont {Z.}~\bibnamefont {Gong}}, \bibinfo {author} {\bibfnamefont {G.-R.}\ \bibnamefont {Yi}}, \bibinfo {author} {\bibfnamefont {S.}~\bibnamefont {Sacanna}},\ and\ \bibinfo {author} {\bibfnamefont {D.~J.}\ \bibnamefont {Pine}},\ }\bibfield  {title} {\bibinfo {title} {Colloidal diamond},\ }\href@noop {} {\bibfield  {journal} {\bibinfo  {journal} {Nature}\ }\textbf {\bibinfo {volume} {585}},\ \bibinfo {pages} {524} (\bibinfo {year} {2020})}\BibitemShut {NoStop}%
\bibitem [{\citenamefont {Yang}\ and\ \citenamefont {Rocchia}(2023)}]{yang2023jpcb}%
  \BibitemOpen
  \bibfield  {author} {\bibinfo {author} {\bibfnamefont {W.}~\bibnamefont {Yang}}\ and\ \bibinfo {author} {\bibfnamefont {W.}~\bibnamefont {Rocchia}},\ }\bibfield  {title} {\bibinfo {title} {Biomolecular electrostatic phenomena: An evergreen field},\ }\href@noop {} {\bibfield  {journal} {\bibinfo  {journal} {The Journal of Physical Chemistry B}\ }\textbf {\bibinfo {volume} {127}},\ \bibinfo {pages} {3979} (\bibinfo {year} {2023})}\BibitemShut {NoStop}%
\bibitem [{\citenamefont {Nakamura}\ and\ \citenamefont {Wada}(1985)}]{nakamura1985nature}%
  \BibitemOpen
  \bibfield  {author} {\bibinfo {author} {\bibfnamefont {H.}~\bibnamefont {Nakamura}}\ and\ \bibinfo {author} {\bibfnamefont {A.}~\bibnamefont {Wada}},\ }\bibfield  {title} {\bibinfo {title} {Nature of the charge distribution in proteins. {III.} electric multipole structures},\ }\href@noop {} {\bibfield  {journal} {\bibinfo  {journal} {Journal of the Physical Society of Japan}\ }\textbf {\bibinfo {volume} {54}},\ \bibinfo {pages} {4047} (\bibinfo {year} {1985})}\BibitemShut {NoStop}%
\bibitem [{\citenamefont {Li}\ \emph {et~al.}(2015)\citenamefont {Li}, \citenamefont {Persson}, \citenamefont {Morin}, \citenamefont {Behrens}, \citenamefont {Lund},\ and\ \citenamefont {Zackrisson~Oskolkova}}]{Li2015Charge}%
  \BibitemOpen
  \bibfield  {author} {\bibinfo {author} {\bibfnamefont {W.}~\bibnamefont {Li}}, \bibinfo {author} {\bibfnamefont {B.~A.}\ \bibnamefont {Persson}}, \bibinfo {author} {\bibfnamefont {M.}~\bibnamefont {Morin}}, \bibinfo {author} {\bibfnamefont {M.~A.}\ \bibnamefont {Behrens}}, \bibinfo {author} {\bibfnamefont {M.}~\bibnamefont {Lund}},\ and\ \bibinfo {author} {\bibfnamefont {M.}~\bibnamefont {Zackrisson~Oskolkova}},\ }\bibfield  {title} {\bibinfo {title} {Charge-induced patchy attractions between proteins},\ }\href@noop {} {\bibfield  {journal} {\bibinfo  {journal} {The Journal of Physical Chemistry B}\ }\textbf {\bibinfo {volume} {119}},\ \bibinfo {pages} {503} (\bibinfo {year} {2015})}\BibitemShut {NoStop}%
\bibitem [{\citenamefont {Lund}(2016)}]{lund2016anisotropic}%
  \BibitemOpen
  \bibfield  {author} {\bibinfo {author} {\bibfnamefont {M.}~\bibnamefont {Lund}},\ }\bibfield  {title} {\bibinfo {title} {Anisotropic protein--protein interactions due to ion binding},\ }\href@noop {} {\bibfield  {journal} {\bibinfo  {journal} {Colloids and Surfaces B: Biointerfaces}\ }\textbf {\bibinfo {volume} {137}},\ \bibinfo {pages} {17} (\bibinfo {year} {2016})}\BibitemShut {NoStop}%
\bibitem [{\citenamefont {{Lošdorfer Božič}}\ and\ \citenamefont {Podgornik}(2017)}]{Bozic2017pH}%
  \BibitemOpen
  \bibfield  {author} {\bibinfo {author} {\bibfnamefont {A.}~\bibnamefont {{Lošdorfer Božič}}}\ and\ \bibinfo {author} {\bibfnamefont {R.}~\bibnamefont {Podgornik}},\ }\bibfield  {title} {\bibinfo {title} {{pH} dependence of charge multipole moments in proteins},\ }\href@noop {} {\bibfield  {journal} {\bibinfo  {journal} {Biophysical Journal}\ }\textbf {\bibinfo {volume} {113}},\ \bibinfo {pages} {1454} (\bibinfo {year} {2017})}\BibitemShut {NoStop}%
\bibitem [{\citenamefont {Boubeta}\ \emph {et~al.}(2018)\citenamefont {Boubeta}, \citenamefont {Soler-Illia},\ and\ \citenamefont {Tagliazucchi}}]{boubeta2018langmuir}%
  \BibitemOpen
  \bibfield  {author} {\bibinfo {author} {\bibfnamefont {F.~M.}\ \bibnamefont {Boubeta}}, \bibinfo {author} {\bibfnamefont {G.~J.~A.~A.}\ \bibnamefont {Soler-Illia}},\ and\ \bibinfo {author} {\bibfnamefont {M.}~\bibnamefont {Tagliazucchi}},\ }\bibfield  {title} {\bibinfo {title} {Electrostatically driven protein adsorption: Charge patches versus charge regulation},\ }\href@noop {} {\bibfield  {journal} {\bibinfo  {journal} {Langmuir}\ }\textbf {\bibinfo {volume} {34}},\ \bibinfo {pages} {15727} (\bibinfo {year} {2018})}\BibitemShut {NoStop}%
\bibitem [{\citenamefont {Lunkad}\ \emph {et~al.}(2022)\citenamefont {Lunkad}, \citenamefont {{Barroso da Silva}},\ and\ \citenamefont {Ko{\v s}ovan}}]{Kosovan_2022}%
  \BibitemOpen
  \bibfield  {author} {\bibinfo {author} {\bibfnamefont {R.}~\bibnamefont {Lunkad}}, \bibinfo {author} {\bibfnamefont {F.~L.}\ \bibnamefont {{Barroso da Silva}}},\ and\ \bibinfo {author} {\bibfnamefont {P.}~\bibnamefont {Ko{\v s}ovan}},\ }\bibfield  {title} {\bibinfo {title} {Both charge-regulation and charge-patch distribution can drive adsorption on the wrong side of the isoelectric point},\ }\href@noop {} {\bibfield  {journal} {\bibinfo  {journal} {J. Am. Chem. Soc.}\ }\textbf {\bibinfo {volume} {144}},\ \bibinfo {pages} {1813} (\bibinfo {year} {2022})}\BibitemShut {NoStop}%
\bibitem [{\citenamefont {Broide}\ \emph {et~al.}(1996)\citenamefont {Broide}, \citenamefont {Tominc},\ and\ \citenamefont {Saxowsky}}]{broide1996using}%
  \BibitemOpen
  \bibfield  {author} {\bibinfo {author} {\bibfnamefont {M.~L.}\ \bibnamefont {Broide}}, \bibinfo {author} {\bibfnamefont {T.~M.}\ \bibnamefont {Tominc}},\ and\ \bibinfo {author} {\bibfnamefont {M.~D.}\ \bibnamefont {Saxowsky}},\ }\bibfield  {title} {\bibinfo {title} {Using phase transitions to investigate the effect of salts on protein interactions},\ }\href {https://doi.org/10.1103/PhysRevE.53.6325} {\bibfield  {journal} {\bibinfo  {journal} {Phys. Rev. E}\ }\textbf {\bibinfo {volume} {53}},\ \bibinfo {pages} {6325} (\bibinfo {year} {1996})}\BibitemShut {NoStop}%
\bibitem [{\citenamefont {Muschol}\ and\ \citenamefont {Rosenberger}(1997)}]{muschol1997liquid}%
  \BibitemOpen
  \bibfield  {author} {\bibinfo {author} {\bibfnamefont {M.}~\bibnamefont {Muschol}}\ and\ \bibinfo {author} {\bibfnamefont {F.}~\bibnamefont {Rosenberger}},\ }\bibfield  {title} {\bibinfo {title} {{Liquid–liquid phase separation in supersaturated lysozyme solutions and associated precipitate formation/crystallization}},\ }\href {https://doi.org/10.1063/1.474547} {\bibfield  {journal} {\bibinfo  {journal} {The Journal of Chemical Physics}\ }\textbf {\bibinfo {volume} {107}},\ \bibinfo {pages} {1953} (\bibinfo {year} {1997})},\ \Eprint {https://arxiv.org/abs/https://pubs.aip.org/aip/jcp/article-pdf/107/6/1953/10788248/1953\_1\_online.pdf} {https://pubs.aip.org/aip/jcp/article-pdf/107/6/1953/10788248/1953\_1\_online.pdf} \BibitemShut {NoStop}%
\bibitem [{\citenamefont {Grigsby}\ \emph {et~al.}(2001)\citenamefont {Grigsby}, \citenamefont {Blanch},\ and\ \citenamefont {Prausnitz}}]{grigsby2001cloud}%
  \BibitemOpen
  \bibfield  {author} {\bibinfo {author} {\bibfnamefont {J.}~\bibnamefont {Grigsby}}, \bibinfo {author} {\bibfnamefont {H.}~\bibnamefont {Blanch}},\ and\ \bibinfo {author} {\bibfnamefont {J.}~\bibnamefont {Prausnitz}},\ }\bibfield  {title} {\bibinfo {title} {Cloud-point temperatures for lysozyme in electrolyte solutions: effect of salt type, salt concentration and ph},\ }\href {https://doi.org/https://doi.org/10.1016/S0301-4622(01)00173-9} {\bibfield  {journal} {\bibinfo  {journal} {Biophysical Chemistry}\ }\textbf {\bibinfo {volume} {91}},\ \bibinfo {pages} {231} (\bibinfo {year} {2001})}\BibitemShut {NoStop}%
\bibitem [{\citenamefont {Gögelein}\ \emph {et~al.}(2012)\citenamefont {Gögelein}, \citenamefont {Wagner}, \citenamefont {Cardinaux}, \citenamefont {Nägele},\ and\ \citenamefont {Egelhaaf}}]{gogelein2012effect}%
  \BibitemOpen
  \bibfield  {author} {\bibinfo {author} {\bibfnamefont {C.}~\bibnamefont {Gögelein}}, \bibinfo {author} {\bibfnamefont {D.}~\bibnamefont {Wagner}}, \bibinfo {author} {\bibfnamefont {F.}~\bibnamefont {Cardinaux}}, \bibinfo {author} {\bibfnamefont {G.}~\bibnamefont {Nägele}},\ and\ \bibinfo {author} {\bibfnamefont {S.~U.}\ \bibnamefont {Egelhaaf}},\ }\bibfield  {title} {\bibinfo {title} {{Effect of glycerol and dimethyl sulfoxide on the phase behavior of lysozyme: Theory and experiments}},\ }\href {https://doi.org/10.1063/1.3673442} {\bibfield  {journal} {\bibinfo  {journal} {The Journal of Chemical Physics}\ }\textbf {\bibinfo {volume} {136}},\ \bibinfo {pages} {015102} (\bibinfo {year} {2012})},\ \Eprint {https://arxiv.org/abs/https://pubs.aip.org/aip/jcp/article-pdf/doi/10.1063/1.3673442/13432171/015102\_1\_online.pdf} {https://pubs.aip.org/aip/jcp/article-pdf/doi/10.1063/1.3673442/13432171/015102\_1\_online.pdf} \BibitemShut {NoStop}%
\bibitem [{\citenamefont {Zhang}\ \emph {et~al.}(2012)\citenamefont {Zhang}, \citenamefont {Roth}, \citenamefont {Wolf}, \citenamefont {Roosen-Runge}, \citenamefont {Skoda}, \citenamefont {Jacobs}, \citenamefont {Stzucki},\ and\ \citenamefont {Schreiber}}]{Zhang2012Charge}%
  \BibitemOpen
  \bibfield  {author} {\bibinfo {author} {\bibfnamefont {F.}~\bibnamefont {Zhang}}, \bibinfo {author} {\bibfnamefont {R.}~\bibnamefont {Roth}}, \bibinfo {author} {\bibfnamefont {M.}~\bibnamefont {Wolf}}, \bibinfo {author} {\bibfnamefont {F.}~\bibnamefont {Roosen-Runge}}, \bibinfo {author} {\bibfnamefont {M.~W.~A.}\ \bibnamefont {Skoda}}, \bibinfo {author} {\bibfnamefont {R.~M.~J.}\ \bibnamefont {Jacobs}}, \bibinfo {author} {\bibfnamefont {M.}~\bibnamefont {Stzucki}},\ and\ \bibinfo {author} {\bibfnamefont {F.}~\bibnamefont {Schreiber}},\ }\bibfield  {title} {\bibinfo {title} {Charge-controlled metastable liquid–liquid phase separation in protein solutions as a universal pathway towards crystallization},\ }\href@noop {} {\bibfield  {journal} {\bibinfo  {journal} {Soft Matter}\ }\textbf {\bibinfo {volume} {8}},\ \bibinfo {pages} {1313} (\bibinfo {year} {2012})}\BibitemShut {NoStop}%
\bibitem [{\citenamefont {Kurut}\ \emph {et~al.}(2012)\citenamefont {Kurut}, \citenamefont {Persson}, \citenamefont {\v{A}kesson}, \citenamefont {Forsman},\ and\ \citenamefont {Lund}}]{lund2012jpcl}%
  \BibitemOpen
  \bibfield  {author} {\bibinfo {author} {\bibfnamefont {A.}~\bibnamefont {Kurut}}, \bibinfo {author} {\bibfnamefont {B.~A.}\ \bibnamefont {Persson}}, \bibinfo {author} {\bibfnamefont {T.}~\bibnamefont {\v{A}kesson}}, \bibinfo {author} {\bibfnamefont {J.}~\bibnamefont {Forsman}},\ and\ \bibinfo {author} {\bibfnamefont {M.}~\bibnamefont {Lund}},\ }\bibfield  {title} {\bibinfo {title} {Anisotropic interactions in protein mixtures: Self assembly and phase behavior in aqueous solution},\ }\href {https://doi.org/10.1021/jz201680m} {\bibfield  {journal} {\bibinfo  {journal} {The Journal of Physical Chemistry Letters}\ }\textbf {\bibinfo {volume} {3}},\ \bibinfo {pages} {731} (\bibinfo {year} {2012})}\BibitemShut {NoStop}%
\bibitem [{\citenamefont {van Oostrum}\ \emph {et~al.}(2015)\citenamefont {van Oostrum}, \citenamefont {Hejazifar}, \citenamefont {Niedermayer},\ and\ \citenamefont {Reimhult}}]{vanostrum2015jpcm}%
  \BibitemOpen
  \bibfield  {author} {\bibinfo {author} {\bibfnamefont {P.~D.~J.}\ \bibnamefont {van Oostrum}}, \bibinfo {author} {\bibfnamefont {M.}~\bibnamefont {Hejazifar}}, \bibinfo {author} {\bibfnamefont {C.}~\bibnamefont {Niedermayer}},\ and\ \bibinfo {author} {\bibfnamefont {E.}~\bibnamefont {Reimhult}},\ }\bibfield  {title} {\bibinfo {title} {Simple method for the synthesis of inverse patchy colloids},\ }\href@noop {} {\bibfield  {journal} {\bibinfo  {journal} {Journal of Physics: Condensed Matter}\ }\textbf {\bibinfo {volume} {27}},\ \bibinfo {pages} {234105} (\bibinfo {year} {2015})}\BibitemShut {NoStop}%
\bibitem [{\citenamefont {Zimmermann}\ \emph {et~al.}(2018)\citenamefont {Zimmermann}, \citenamefont {Grigoriev}, \citenamefont {Puretskiy},\ and\ \citenamefont {B{\"o}ker}}]{Zimmermann_2018}%
  \BibitemOpen
  \bibfield  {author} {\bibinfo {author} {\bibfnamefont {M.}~\bibnamefont {Zimmermann}}, \bibinfo {author} {\bibfnamefont {D.}~\bibnamefont {Grigoriev}}, \bibinfo {author} {\bibfnamefont {N.}~\bibnamefont {Puretskiy}},\ and\ \bibinfo {author} {\bibfnamefont {A.}~\bibnamefont {B{\"o}ker}},\ }\bibfield  {title} {\bibinfo {title} {Characteristics of microcontact printing with polyelectrolyte ink for the precise preparation of patches on silica particles},\ }\href@noop {} {\bibfield  {journal} {\bibinfo  {journal} {RSC Adv.}\ }\textbf {\bibinfo {volume} {8}},\ \bibinfo {pages} {39241} (\bibinfo {year} {2018})}\BibitemShut {NoStop}%
\bibitem [{\citenamefont {Mehr}\ \emph {et~al.}(2019)\citenamefont {Mehr}, \citenamefont {Grigoriev}, \citenamefont {Puretskiy},\ and\ \citenamefont {B{\"o}ker}}]{Mehr2019sm}%
  \BibitemOpen
  \bibfield  {author} {\bibinfo {author} {\bibfnamefont {F.~N.}\ \bibnamefont {Mehr}}, \bibinfo {author} {\bibfnamefont {D.}~\bibnamefont {Grigoriev}}, \bibinfo {author} {\bibfnamefont {N.}~\bibnamefont {Puretskiy}},\ and\ \bibinfo {author} {\bibfnamefont {A.}~\bibnamefont {B{\"o}ker}},\ }\bibfield  {title} {\bibinfo {title} {Mono-patchy zwitterionic microcolloids as building blocks for ph-controlled self-assembly},\ }\href@noop {} {\bibfield  {journal} {\bibinfo  {journal} {Soft Matter}\ }\textbf {\bibinfo {volume} {15}},\ \bibinfo {pages} {2430} (\bibinfo {year} {2019})}\BibitemShut {NoStop}%
\bibitem [{\citenamefont {Shanmugathasan}\ \emph {et~al.}(2022)\citenamefont {Shanmugathasan}, \citenamefont {Bagur}, \citenamefont {Ducrot}, \citenamefont {Buffière}, \citenamefont {{van Oostrum}}, \citenamefont {Ravaine},\ and\ \citenamefont {Duguet}}]{Shanmugathasan2022Silica}%
  \BibitemOpen
  \bibfield  {author} {\bibinfo {author} {\bibfnamefont {S.}~\bibnamefont {Shanmugathasan}}, \bibinfo {author} {\bibfnamefont {A.}~\bibnamefont {Bagur}}, \bibinfo {author} {\bibfnamefont {E.}~\bibnamefont {Ducrot}}, \bibinfo {author} {\bibfnamefont {S.}~\bibnamefont {Buffière}}, \bibinfo {author} {\bibfnamefont {P.}~\bibnamefont {{van Oostrum}}}, \bibinfo {author} {\bibfnamefont {S.}~\bibnamefont {Ravaine}},\ and\ \bibinfo {author} {\bibfnamefont {E.}~\bibnamefont {Duguet}},\ }\bibfield  {title} {\bibinfo {title} {Silica/polystyrene bipod-like submicron colloids synthesized by seed-growth dispersion polymerisation as precursors for two-patch silica particles},\ }\href {https://doi.org/https://doi.org/10.1016/j.colsurfa.2022.129344} {\bibfield  {journal} {\bibinfo  {journal} {Colloids and Surfaces A: Physicochemical and Engineering Aspects}\ }\textbf {\bibinfo {volume} {648}},\ \bibinfo {pages} {129344} (\bibinfo {year} {2022})}\BibitemShut {NoStop}%
\bibitem [{\citenamefont {Virk}\ \emph {et~al.}(2023)\citenamefont {Virk}, \citenamefont {Beitl},\ and\ \citenamefont {van Oostrum}}]{Virk2023Synthesis}%
  \BibitemOpen
  \bibfield  {author} {\bibinfo {author} {\bibfnamefont {M.~M.}\ \bibnamefont {Virk}}, \bibinfo {author} {\bibfnamefont {K.~N.}\ \bibnamefont {Beitl}},\ and\ \bibinfo {author} {\bibfnamefont {P.~D.~J.}\ \bibnamefont {van Oostrum}},\ }\bibfield  {title} {\bibinfo {title} {Synthesis of patchy particles using gaseous ligands},\ }\href {https://doi.org/10.1088/1361-648X/acbddc} {\bibfield  {journal} {\bibinfo  {journal} {Journal of Physics: Condensed Matter}\ }\textbf {\bibinfo {volume} {35}},\ \bibinfo {pages} {174003} (\bibinfo {year} {2023})}\BibitemShut {NoStop}%
\bibitem [{\citenamefont {Sabapathy}\ \emph {et~al.}(2017)\citenamefont {Sabapathy}, \citenamefont {Mathews},\ and\ \citenamefont {Mani}}]{sabapathy2017pccp}%
  \BibitemOpen
  \bibfield  {author} {\bibinfo {author} {\bibfnamefont {M.}~\bibnamefont {Sabapathy}}, \bibinfo {author} {\bibfnamefont {R.~A.}\ \bibnamefont {Mathews}},\ and\ \bibinfo {author} {\bibfnamefont {E.}~\bibnamefont {Mani}},\ }\bibfield  {title} {\bibinfo {title} {Self-assembly of inverse patchy colloids with tunable patch coverage},\ }\href@noop {} {\bibfield  {journal} {\bibinfo  {journal} {Physical Chemistry Chemical Physics}\ }\textbf {\bibinfo {volume} {19}},\ \bibinfo {pages} {13122} (\bibinfo {year} {2017})}\BibitemShut {NoStop}%
\bibitem [{\citenamefont {Noguchi}\ \emph {et~al.}(2019)\citenamefont {Noguchi}, \citenamefont {Iwashita},\ and\ \citenamefont {Kimura}}]{Kimura_2019}%
  \BibitemOpen
  \bibfield  {author} {\bibinfo {author} {\bibfnamefont {T.~G.}\ \bibnamefont {Noguchi}}, \bibinfo {author} {\bibfnamefont {Y.}~\bibnamefont {Iwashita}},\ and\ \bibinfo {author} {\bibfnamefont {Y.}~\bibnamefont {Kimura}},\ }\bibfield  {title} {\bibinfo {title} {Controlled armoring of metal surfaces with metallodielectric patchy particles},\ }\href@noop {} {\bibfield  {journal} {\bibinfo  {journal} {J. Chem. Phys.}\ }\textbf {\bibinfo {volume} {150}},\ \bibinfo {pages} {174903} (\bibinfo {year} {2019})}\BibitemShut {NoStop}%
\bibitem [{\citenamefont {Naderi~Mehr}\ \emph {et~al.}(2020)\citenamefont {Naderi~Mehr}, \citenamefont {Grigoriev}, \citenamefont {Heaton}, \citenamefont {Baptiste}, \citenamefont {Stace}, \citenamefont {Puretskiy}, \citenamefont {Besley},\ and\ \citenamefont {Böker}}]{Naderi2020Self}%
  \BibitemOpen
  \bibfield  {author} {\bibinfo {author} {\bibfnamefont {F.}~\bibnamefont {Naderi~Mehr}}, \bibinfo {author} {\bibfnamefont {D.}~\bibnamefont {Grigoriev}}, \bibinfo {author} {\bibfnamefont {R.}~\bibnamefont {Heaton}}, \bibinfo {author} {\bibfnamefont {J.}~\bibnamefont {Baptiste}}, \bibinfo {author} {\bibfnamefont {A.~J.}\ \bibnamefont {Stace}}, \bibinfo {author} {\bibfnamefont {N.}~\bibnamefont {Puretskiy}}, \bibinfo {author} {\bibfnamefont {E.}~\bibnamefont {Besley}},\ and\ \bibinfo {author} {\bibfnamefont {A.}~\bibnamefont {Böker}},\ }\bibfield  {title} {\bibinfo {title} {Self-assembly behavior of oppositely charged inverse bipatchy microcolloids},\ }\href@noop {} {\bibfield  {journal} {\bibinfo  {journal} {Small}\ }\textbf {\bibinfo {volume} {16}},\ \bibinfo {pages} {2000442} (\bibinfo {year} {2020})}\BibitemShut {NoStop}%
\bibitem [{\citenamefont {Lebdioua}\ \emph {et~al.}(2021)\citenamefont {Lebdioua}, \citenamefont {Cerbelaud}, \citenamefont {Aimable},\ and\ \citenamefont {Videcoq}}]{Lebdioua2021Study}%
  \BibitemOpen
  \bibfield  {author} {\bibinfo {author} {\bibfnamefont {K.}~\bibnamefont {Lebdioua}}, \bibinfo {author} {\bibfnamefont {M.}~\bibnamefont {Cerbelaud}}, \bibinfo {author} {\bibfnamefont {A.}~\bibnamefont {Aimable}},\ and\ \bibinfo {author} {\bibfnamefont {A.}~\bibnamefont {Videcoq}},\ }\bibfield  {title} {\bibinfo {title} {Study of the aggregation behavior of janus particles by coupling experiments and brownian dynamics simulations},\ }\href {https://doi.org/https://doi.org/10.1016/j.jcis.2020.09.031} {\bibfield  {journal} {\bibinfo  {journal} {Journal of Colloid and Interface Science}\ }\textbf {\bibinfo {volume} {583}},\ \bibinfo {pages} {222} (\bibinfo {year} {2021})}\BibitemShut {NoStop}%
\bibitem [{\citenamefont {Dempster}\ and\ \citenamefont {{de la Cruz}}(2016)}]{Cruz_2016}%
  \BibitemOpen
  \bibfield  {author} {\bibinfo {author} {\bibfnamefont {J.~M.}\ \bibnamefont {Dempster}}\ and\ \bibinfo {author} {\bibfnamefont {M.~O.}\ \bibnamefont {{de la Cruz}}},\ }\bibfield  {title} {\bibinfo {title} {Aggregation of heterogeneously charged colloids},\ }\href@noop {} {\bibfield  {journal} {\bibinfo  {journal} {ACS Nano}\ }\textbf {\bibinfo {volume} {10}},\ \bibinfo {pages} {5909} (\bibinfo {year} {2016})}\BibitemShut {NoStop}%
\bibitem [{\citenamefont {Blanco}\ and\ \citenamefont {Shen}(2016)}]{Blanco_2016}%
  \BibitemOpen
  \bibfield  {author} {\bibinfo {author} {\bibfnamefont {M.~A.}\ \bibnamefont {Blanco}}\ and\ \bibinfo {author} {\bibfnamefont {V.~K.}\ \bibnamefont {Shen}},\ }\bibfield  {title} {\bibinfo {title} {Effect of the surface charge distribution on the fluid phase behavior of charged colloids and proteins},\ }\href@noop {} {\bibfield  {journal} {\bibinfo  {journal} {J. Chem. Phys.}\ }\textbf {\bibinfo {volume} {145}},\ \bibinfo {pages} {155102} (\bibinfo {year} {2016})}\BibitemShut {NoStop}%
\bibitem [{\citenamefont {Ferrari}\ \emph {et~al.}(2017)\citenamefont {Ferrari}, \citenamefont {Bianchi},\ and\ \citenamefont {Kahl}}]{silvanonanoscale}%
  \BibitemOpen
  \bibfield  {author} {\bibinfo {author} {\bibfnamefont {S.}~\bibnamefont {Ferrari}}, \bibinfo {author} {\bibfnamefont {E.}~\bibnamefont {Bianchi}},\ and\ \bibinfo {author} {\bibfnamefont {G.}~\bibnamefont {Kahl}},\ }\bibfield  {title} {\bibinfo {title} {Spontaneous assembly of a hybrid crystal-liquid phase in inverse patchy colloid systems},\ }\href@noop {} {\bibfield  {journal} {\bibinfo  {journal} {Nanoscale}\ }\textbf {\bibinfo {volume} {9}},\ \bibinfo {pages} {1956} (\bibinfo {year} {2017})}\BibitemShut {NoStop}%
\bibitem [{\citenamefont {Abrikosov}\ \emph {et~al.}(2017)\citenamefont {Abrikosov}, \citenamefont {Stenqvist},\ and\ \citenamefont {Lund}}]{abrikosov2017steering}%
  \BibitemOpen
  \bibfield  {author} {\bibinfo {author} {\bibfnamefont {A.~I.}\ \bibnamefont {Abrikosov}}, \bibinfo {author} {\bibfnamefont {B.}~\bibnamefont {Stenqvist}},\ and\ \bibinfo {author} {\bibfnamefont {M.}~\bibnamefont {Lund}},\ }\bibfield  {title} {\bibinfo {title} {Steering patchy particles using multivalent electrolytes},\ }\href@noop {} {\bibfield  {journal} {\bibinfo  {journal} {Soft matter}\ }\textbf {\bibinfo {volume} {13}},\ \bibinfo {pages} {4591} (\bibinfo {year} {2017})}\BibitemShut {NoStop}%
\bibitem [{\citenamefont {{de Ara{\'u}jo}}\ \emph {et~al.}(2017)\citenamefont {{de Ara{\'u}jo}}, \citenamefont {Munarin}, \citenamefont {Farias}, \citenamefont {Peeters},\ and\ \citenamefont {Ferreira}}]{Ferreira_2017}%
  \BibitemOpen
  \bibfield  {author} {\bibinfo {author} {\bibfnamefont {J.~L.~B.}\ \bibnamefont {{de Ara{\'u}jo}}}, \bibinfo {author} {\bibfnamefont {F.~F.}\ \bibnamefont {Munarin}}, \bibinfo {author} {\bibfnamefont {G.~A.}\ \bibnamefont {Farias}}, \bibinfo {author} {\bibfnamefont {F.~M.}\ \bibnamefont {Peeters}},\ and\ \bibinfo {author} {\bibfnamefont {W.~P.}\ \bibnamefont {Ferreira}},\ }\bibfield  {title} {\bibinfo {title} {Structure and reentrant percolation in an inverse patchy colloidal system},\ }\href@noop {} {\bibfield  {journal} {\bibinfo  {journal} {Phys. Rev. E}\ }\textbf {\bibinfo {volume} {95}},\ \bibinfo {pages} {062606} (\bibinfo {year} {2017})}\BibitemShut {NoStop}%
\bibitem [{\citenamefont {Cerbelaud}\ \emph {et~al.}(2019)\citenamefont {Cerbelaud}, \citenamefont {Lebdioua}, \citenamefont {{Tam Tran}}, \citenamefont {Crespin}, \citenamefont {Aimable},\ and\ \citenamefont {Videcoq}}]{Cerbelaud_2019}%
  \BibitemOpen
  \bibfield  {author} {\bibinfo {author} {\bibfnamefont {M.}~\bibnamefont {Cerbelaud}}, \bibinfo {author} {\bibfnamefont {K.}~\bibnamefont {Lebdioua}}, \bibinfo {author} {\bibfnamefont {C.}~\bibnamefont {{Tam Tran}}}, \bibinfo {author} {\bibfnamefont {B.}~\bibnamefont {Crespin}}, \bibinfo {author} {\bibfnamefont {A.}~\bibnamefont {Aimable}},\ and\ \bibinfo {author} {\bibfnamefont {A.}~\bibnamefont {Videcoq}},\ }\bibfield  {title} {\bibinfo {title} {Brownian dynamics simulations of one-patch inverse patchy particles},\ }\href@noop {} {\bibfield  {journal} {\bibinfo  {journal} {Phys.Chem.Chem.Phys.}\ }\textbf {\bibinfo {volume} {21}},\ \bibinfo {pages} {23447} (\bibinfo {year} {2019})}\BibitemShut {NoStop}%
\bibitem [{\citenamefont {Wang}\ and\ \citenamefont {Swan}(2019)}]{Swan_2019}%
  \BibitemOpen
  \bibfield  {author} {\bibinfo {author} {\bibfnamefont {G.}~\bibnamefont {Wang}}\ and\ \bibinfo {author} {\bibfnamefont {J.~W.}\ \bibnamefont {Swan}},\ }\bibfield  {title} {\bibinfo {title} {Surface heterogeneity affects percolation and gelation of colloids: dynamic simulations with random patchy spheres},\ }\href@noop {} {\bibfield  {journal} {\bibinfo  {journal} {Soft Matter}\ }\textbf {\bibinfo {volume} {215}},\ \bibinfo {pages} {5094} (\bibinfo {year} {2019})}\BibitemShut {NoStop}%
\bibitem [{\citenamefont {Rocha}\ \emph {et~al.}(2021)\citenamefont {Rocha}, \citenamefont {Paul},\ and\ \citenamefont {Vashisth}}]{Vashisth_2021}%
  \BibitemOpen
  \bibfield  {author} {\bibinfo {author} {\bibfnamefont {B.~C.}\ \bibnamefont {Rocha}}, \bibinfo {author} {\bibfnamefont {S.}~\bibnamefont {Paul}},\ and\ \bibinfo {author} {\bibfnamefont {H.}~\bibnamefont {Vashisth}},\ }\bibfield  {title} {\bibinfo {title} {Enhanced porosity in self-assembled morphologies mediated by charged lobes on patchy particles},\ }\href@noop {} {\bibfield  {journal} {\bibinfo  {journal} {J. Phys. Chem. B}\ }\textbf {\bibinfo {volume} {125}},\ \bibinfo {pages} {3208} (\bibinfo {year} {2021})}\BibitemShut {NoStop}%
\bibitem [{\citenamefont {K}\ and\ \citenamefont {Mani}(2021)}]{mani2021stabilizing}%
  \BibitemOpen
  \bibfield  {author} {\bibinfo {author} {\bibfnamefont {R.~A.~M.}\ \bibnamefont {K}}\ and\ \bibinfo {author} {\bibfnamefont {E.}~\bibnamefont {Mani}},\ }\bibfield  {title} {\bibinfo {title} {Stabilizing ordered structures with single patch inverse patchy colloids in two dimensions},\ }\href@noop {} {\bibfield  {journal} {\bibinfo  {journal} {Journal of Physics: Condensed Matter}\ }\textbf {\bibinfo {volume} {33}},\ \bibinfo {pages} {195101} (\bibinfo {year} {2021})}\BibitemShut {NoStop}%
\bibitem [{\citenamefont {Hoffmann}\ \emph {et~al.}(2004)\citenamefont {Hoffmann}, \citenamefont {Likos},\ and\ \citenamefont {Hansen}}]{hoffmann2004molphys}%
  \BibitemOpen
  \bibfield  {author} {\bibinfo {author} {\bibfnamefont {N.}~\bibnamefont {Hoffmann}}, \bibinfo {author} {\bibfnamefont {C.~N.}\ \bibnamefont {Likos}},\ and\ \bibinfo {author} {\bibfnamefont {J.-P.}\ \bibnamefont {Hansen}},\ }\bibfield  {title} {\bibinfo {title} {Linear screening of the electrostatic potential around spherical particles with non-spherical charge patterns},\ }\href@noop {} {\bibfield  {journal} {\bibinfo  {journal} {Molecular Physics}\ }\textbf {\bibinfo {volume} {102}},\ \bibinfo {pages} {857} (\bibinfo {year} {2004})}\BibitemShut {NoStop}%
\bibitem [{\citenamefont {Boon}\ \emph {et~al.}(2010)\citenamefont {Boon}, \citenamefont {{Carvajal Gallardo}}, \citenamefont {Zheng}, \citenamefont {Eggen}, \citenamefont {Dijkstra},\ and\ \citenamefont {{van Roij}}}]{boon2010jpcm}%
  \BibitemOpen
  \bibfield  {author} {\bibinfo {author} {\bibfnamefont {N.}~\bibnamefont {Boon}}, \bibinfo {author} {\bibfnamefont {E.}~\bibnamefont {{Carvajal Gallardo}}}, \bibinfo {author} {\bibfnamefont {S.}~\bibnamefont {Zheng}}, \bibinfo {author} {\bibfnamefont {E.}~\bibnamefont {Eggen}}, \bibinfo {author} {\bibfnamefont {M.}~\bibnamefont {Dijkstra}},\ and\ \bibinfo {author} {\bibfnamefont {R.}~\bibnamefont {{van Roij}}},\ }\bibfield  {title} {\bibinfo {title} {Screening of heterogeneous surfaces: charge renormalization of {Janus} particles},\ }\href@noop {} {\bibfield  {journal} {\bibinfo  {journal} {Journal of Physics: Condensed Matter}\ }\textbf {\bibinfo {volume} {22}},\ \bibinfo {pages} {104104} (\bibinfo {year} {2010})}\BibitemShut {NoStop}%
\bibitem [{\citenamefont {Bianchi}\ \emph {et~al.}(2011{\natexlab{b}})\citenamefont {Bianchi}, \citenamefont {Kahl},\ and\ \citenamefont {Likos}}]{Bianchi2011Inverse}%
  \BibitemOpen
  \bibfield  {author} {\bibinfo {author} {\bibfnamefont {E.}~\bibnamefont {Bianchi}}, \bibinfo {author} {\bibfnamefont {G.}~\bibnamefont {Kahl}},\ and\ \bibinfo {author} {\bibfnamefont {C.~N.}\ \bibnamefont {Likos}},\ }\bibfield  {title} {\bibinfo {title} {Inverse patchy colloids: from microscopic description to mesoscopic coarse-graining},\ }\href {https://doi.org/10.1039/C1SM05597F} {\bibfield  {journal} {\bibinfo  {journal} {Soft Matter}\ }\textbf {\bibinfo {volume} {7}},\ \bibinfo {pages} {8313} (\bibinfo {year} {2011}{\natexlab{b}})}\BibitemShut {NoStop}%
\bibitem [{\citenamefont {{de Graaf}}\ \emph {et~al.}(2012)\citenamefont {{de Graaf}}, \citenamefont {Boon}, \citenamefont {Dijkstra},\ and\ \citenamefont {{van Roij}}}]{degraaf2012jcp}%
  \BibitemOpen
  \bibfield  {author} {\bibinfo {author} {\bibfnamefont {J.}~\bibnamefont {{de Graaf}}}, \bibinfo {author} {\bibfnamefont {N.}~\bibnamefont {Boon}}, \bibinfo {author} {\bibfnamefont {M.}~\bibnamefont {Dijkstra}},\ and\ \bibinfo {author} {\bibfnamefont {R.}~\bibnamefont {{van Roij}}},\ }\bibfield  {title} {\bibinfo {title} {Electrostatic interactions between {Janus} particles},\ }\href@noop {} {\bibfield  {journal} {\bibinfo  {journal} {Journal of Chemical Physics}\ }\textbf {\bibinfo {volume} {137}},\ \bibinfo {pages} {104910} (\bibinfo {year} {2012})}\BibitemShut {NoStop}%
\bibitem [{\citenamefont {Yigit}\ \emph {et~al.}(2015)\citenamefont {Yigit}, \citenamefont {Heyda},\ and\ \citenamefont {Dzubiella}}]{yigit2015jcp}%
  \BibitemOpen
  \bibfield  {author} {\bibinfo {author} {\bibfnamefont {C.}~\bibnamefont {Yigit}}, \bibinfo {author} {\bibfnamefont {J.}~\bibnamefont {Heyda}},\ and\ \bibinfo {author} {\bibfnamefont {J.}~\bibnamefont {Dzubiella}},\ }\bibfield  {title} {\bibinfo {title} {Charged patchy particle models in explicit salt: ion distributions, electrostatic potentials, and effective interactions},\ }\href@noop {} {\bibfield  {journal} {\bibinfo  {journal} {The Journal of Chemical Physics}\ }\textbf {\bibinfo {volume} {143}},\ \bibinfo {pages} {064904} (\bibinfo {year} {2015})}\BibitemShut {NoStop}%
\bibitem [{\citenamefont {Hieronimus}\ \emph {et~al.}(2016)\citenamefont {Hieronimus}, \citenamefont {Raschke},\ and\ \citenamefont {Heuer}}]{hieronimus2016jcp}%
  \BibitemOpen
  \bibfield  {author} {\bibinfo {author} {\bibfnamefont {R.}~\bibnamefont {Hieronimus}}, \bibinfo {author} {\bibfnamefont {S.}~\bibnamefont {Raschke}},\ and\ \bibinfo {author} {\bibfnamefont {A.}~\bibnamefont {Heuer}},\ }\bibfield  {title} {\bibinfo {title} {How to model the interaction of charged {J}anus particles},\ }\href@noop {} {\bibfield  {journal} {\bibinfo  {journal} {The Journal of Chemical Physics}\ }\textbf {\bibinfo {volume} {145}},\ \bibinfo {pages} {064303} (\bibinfo {year} {2016})}\BibitemShut {NoStop}%
\bibitem [{\citenamefont {Brunk}\ \emph {et~al.}(2020)\citenamefont {Brunk}, \citenamefont {Kadupitiya},\ and\ \citenamefont {Jadhao}}]{Brunk_2020}%
  \BibitemOpen
  \bibfield  {author} {\bibinfo {author} {\bibfnamefont {N.~E.}\ \bibnamefont {Brunk}}, \bibinfo {author} {\bibfnamefont {J.}~\bibnamefont {Kadupitiya}},\ and\ \bibinfo {author} {\bibfnamefont {V.}~\bibnamefont {Jadhao}},\ }\bibfield  {title} {\bibinfo {title} {Designing surface charge patterns for shape control of deformable nanoparticles},\ }\href@noop {} {\bibfield  {journal} {\bibinfo  {journal} {Physical Review Letters}\ }\textbf {\bibinfo {volume} {125}},\ \bibinfo {pages} {248001} (\bibinfo {year} {2020})}\BibitemShut {NoStop}%
\bibitem [{\citenamefont {Mathews~Kalapurakal}\ and\ \citenamefont {Mani}(2022)}]{mathews2022molsim}%
  \BibitemOpen
  \bibfield  {author} {\bibinfo {author} {\bibfnamefont {R.~A.}\ \bibnamefont {Mathews~Kalapurakal}}\ and\ \bibinfo {author} {\bibfnamefont {E.}~\bibnamefont {Mani}},\ }\bibfield  {title} {\bibinfo {title} {Orientation-dependent electrostatic interaction between inverse patchy colloids},\ }\href@noop {} {\bibfield  {journal} {\bibinfo  {journal} {Molecular Simulation}\ }\textbf {\bibinfo {volume} {48}},\ \bibinfo {pages} {176} (\bibinfo {year} {2022})}\BibitemShut {NoStop}%
\bibitem [{\citenamefont {Popov}\ and\ \citenamefont {Hernandez}(2023)}]{popov2023jpcb}%
  \BibitemOpen
  \bibfield  {author} {\bibinfo {author} {\bibfnamefont {A.}~\bibnamefont {Popov}}\ and\ \bibinfo {author} {\bibfnamefont {R.}~\bibnamefont {Hernandez}},\ }\bibfield  {title} {\bibinfo {title} {Bottom-up construction of the interaction between {Janus} particles},\ }\href@noop {} {\bibfield  {journal} {\bibinfo  {journal} {The Journal of Physical Chemistry B}\ }\textbf {\bibinfo {volume} {127}},\ \bibinfo {pages} {1664} (\bibinfo {year} {2023})}\BibitemShut {NoStop}%
\bibitem [{\citenamefont {Stipsitz}\ \emph {et~al.}(2015)\citenamefont {Stipsitz}, \citenamefont {Bianchi},\ and\ \citenamefont {Kahl}}]{bianchi:2015}%
  \BibitemOpen
  \bibfield  {author} {\bibinfo {author} {\bibfnamefont {M.}~\bibnamefont {Stipsitz}}, \bibinfo {author} {\bibfnamefont {E.}~\bibnamefont {Bianchi}},\ and\ \bibinfo {author} {\bibfnamefont {G.}~\bibnamefont {Kahl}},\ }\bibfield  {title} {\bibinfo {title} {Generalized inverse patchy colloid model},\ }\href@noop {} {\bibfield  {journal} {\bibinfo  {journal} {J. Chem. Phys.}\ }\textbf {\bibinfo {volume} {142}},\ \bibinfo {pages} {114905} (\bibinfo {year} {2015})}\BibitemShut {NoStop}%
\bibitem [{\citenamefont {Noya}\ \emph {et~al.}(2014)\citenamefont {Noya}, \citenamefont {Kolovos}, \citenamefont {Doppelbauer}, \citenamefont {Kahl},\ and\ \citenamefont {Bianchi}}]{Noya2014}%
  \BibitemOpen
  \bibfield  {author} {\bibinfo {author} {\bibfnamefont {E.~G.}\ \bibnamefont {Noya}}, \bibinfo {author} {\bibfnamefont {I.}~\bibnamefont {Kolovos}}, \bibinfo {author} {\bibfnamefont {G.}~\bibnamefont {Doppelbauer}}, \bibinfo {author} {\bibfnamefont {G.}~\bibnamefont {Kahl}},\ and\ \bibinfo {author} {\bibfnamefont {E.}~\bibnamefont {Bianchi}},\ }\bibfield  {title} {\bibinfo {title} {Phase diagram of inverse patchy colloids assembling into an equilibrium laminar phase},\ }\href@noop {} {\bibfield  {journal} {\bibinfo  {journal} {Soft Matter}\ }\textbf {\bibinfo {volume} {10}},\ \bibinfo {pages} {8464} (\bibinfo {year} {2014})}\BibitemShut {NoStop}%
\bibitem [{\citenamefont {Rovigatti}\ \emph {et~al.}(2018)\citenamefont {Rovigatti}, \citenamefont {Russo},\ and\ \citenamefont {Romano}}]{Rovigatti2018How}%
  \BibitemOpen
  \bibfield  {author} {\bibinfo {author} {\bibfnamefont {L.}~\bibnamefont {Rovigatti}}, \bibinfo {author} {\bibfnamefont {J.}~\bibnamefont {Russo}},\ and\ \bibinfo {author} {\bibfnamefont {F.}~\bibnamefont {Romano}},\ }\bibfield  {title} {\bibinfo {title} {How to simulate patchy particles},\ }\bibfield  {journal} {\bibinfo  {journal} {The European Physical Journal E}\ }\textbf {\bibinfo {volume} {41}},\ \href {https://doi.org/10.1140/epje/i2018-11667-x} {10.1140/epje/i2018-11667-x} (\bibinfo {year} {2018})\BibitemShut {NoStop}%
\bibitem [{\citenamefont {Ferrenberg}\ and\ \citenamefont {Swendsen}(1988)}]{Ferrenberg1988New}%
  \BibitemOpen
  \bibfield  {author} {\bibinfo {author} {\bibfnamefont {A.~M.}\ \bibnamefont {Ferrenberg}}\ and\ \bibinfo {author} {\bibfnamefont {R.~H.}\ \bibnamefont {Swendsen}},\ }\bibfield  {title} {\bibinfo {title} {New monte carlo technique for studying phase transitions},\ }\href {https://doi.org/10.1103/PhysRevLett.61.2635} {\bibfield  {journal} {\bibinfo  {journal} {Phys. Rev. Lett.}\ }\textbf {\bibinfo {volume} {61}},\ \bibinfo {pages} {2635} (\bibinfo {year} {1988})}\BibitemShut {NoStop}%
\bibitem [{\citenamefont {Bruce}\ and\ \citenamefont {Wilding}(1992)}]{Bruce1992Scaling}%
  \BibitemOpen
  \bibfield  {author} {\bibinfo {author} {\bibfnamefont {A.~D.}\ \bibnamefont {Bruce}}\ and\ \bibinfo {author} {\bibfnamefont {N.~B.}\ \bibnamefont {Wilding}},\ }\bibfield  {title} {\bibinfo {title} {Scaling fields and universality of the liquid-gas critical point},\ }\href {https://doi.org/10.1103/PhysRevLett.68.193} {\bibfield  {journal} {\bibinfo  {journal} {Phys. Rev. Lett.}\ }\textbf {\bibinfo {volume} {68}},\ \bibinfo {pages} {193} (\bibinfo {year} {1992})}\BibitemShut {NoStop}%
\bibitem [{\citenamefont {Foffi}\ and\ \citenamefont {Sciortino}(2007)}]{Foffi2007}%
  \BibitemOpen
  \bibfield  {author} {\bibinfo {author} {\bibfnamefont {G.}~\bibnamefont {Foffi}}\ and\ \bibinfo {author} {\bibfnamefont {F.}~\bibnamefont {Sciortino}},\ }\bibfield  {title} {\bibinfo {title} {On the possibility of extending the noro-frenkel generalized law of correspondent states to nonisotropic patchy interactions},\ }\href@noop {} {\bibfield  {journal} {\bibinfo  {journal} {J. Phys. Chem. B}\ }\textbf {\bibinfo {volume} {111}},\ \bibinfo {pages} {9702} (\bibinfo {year} {2007})}\BibitemShut {NoStop}%
\bibitem [{\citenamefont {Stauffer}\ and\ \citenamefont {Aharony}(1992)}]{Stauffer1992Introduction}%
  \BibitemOpen
  \bibfield  {author} {\bibinfo {author} {\bibfnamefont {D.}~\bibnamefont {Stauffer}}\ and\ \bibinfo {author} {\bibfnamefont {A.}~\bibnamefont {Aharony}},\ }\href@noop {} {\emph {\bibinfo {title} {Introduction To Percolation Theory: Second Edition (2nd ed.)}}}\ (\bibinfo  {publisher} {Taylor \& Francis},\ \bibinfo {address} {London},\ \bibinfo {year} {1992})\BibitemShut {NoStop}%
\bibitem [{\citenamefont {Notarmuzi}\ and\ \citenamefont {Bianchi}()}]{unpublished}%
  \BibitemOpen
  \bibfield  {author} {\bibinfo {author} {\bibfnamefont {D.}~\bibnamefont {Notarmuzi}}\ and\ \bibinfo {author} {\bibfnamefont {E.}~\bibnamefont {Bianchi}},\ }\href@noop {} {\bibinfo  {journal} {In preparation}\ }\BibitemShut {NoStop}%
\bibitem [{\citenamefont {Tsypin}\ and\ \citenamefont {Bl\"ote}(2000)}]{Tsypin2000Probability}%
  \BibitemOpen
\bibfield  {journal} {  }\bibfield  {author} {\bibinfo {author} {\bibfnamefont {M.~M.}\ \bibnamefont {Tsypin}}\ and\ \bibinfo {author} {\bibfnamefont {H.~W.~J.}\ \bibnamefont {Bl\"ote}},\ }\bibfield  {title} {\bibinfo {title} {Probability distribution of the order parameter for the three-dimensional ising-model universality class: A high-precision monte carlo study},\ }\href {https://doi.org/10.1103/PhysRevE.62.73} {\bibfield  {journal} {\bibinfo  {journal} {Phys. Rev. E}\ }\textbf {\bibinfo {volume} {62}},\ \bibinfo {pages} {73} (\bibinfo {year} {2000})}\BibitemShut {NoStop}%
\bibitem [{\citenamefont {Sciortino}\ \emph {et~al.}(2007)\citenamefont {Sciortino}, \citenamefont {Bianchi}, \citenamefont {Douglas},\ and\ \citenamefont {Tartaglia}}]{Sciortino2007Self}%
  \BibitemOpen
  \bibfield  {author} {\bibinfo {author} {\bibfnamefont {F.}~\bibnamefont {Sciortino}}, \bibinfo {author} {\bibfnamefont {E.}~\bibnamefont {Bianchi}}, \bibinfo {author} {\bibfnamefont {J.~F.}\ \bibnamefont {Douglas}},\ and\ \bibinfo {author} {\bibfnamefont {P.}~\bibnamefont {Tartaglia}},\ }\bibfield  {title} {\bibinfo {title} {{Self-assembly of patchy particles into polymer chains: A parameter-free comparison between Wertheim theory and Monte Carlo simulation}},\ }\href {https://doi.org/10.1063/1.2730797} {\bibfield  {journal} {\bibinfo  {journal} {The Journal of Chemical Physics}\ }\textbf {\bibinfo {volume} {126}},\ \bibinfo {pages} {194903} (\bibinfo {year} {2007})}\BibitemShut {NoStop}%
\bibitem [{Git()}]{GitCode}%
  \BibitemOpen
  \href@noop {} {\bibinfo {title} {Code to reproduce the results}},\ \bibinfo {howpublished} {\url{https://github.com/DaniMuzi/IPPs-critical-point}}\BibitemShut {NoStop}%
\end{thebibliography}

% THE LINE ABOVE IS REPLACED BY THE CONTENT OF THE .bbl FILE.
% THE CONTENT OF THE .bbl FILE FOLLOWS


\end{document}